\pdfoutput=1
\documentclass{midl}

\usepackage{booktabs}
\usepackage{array}
\usepackage{placeins}
\usepackage{makecell}
\usepackage{multirow}
\usepackage{xcolor}

\jmlrproceedings{}{Preprint}
\jmlrvolume{}
\jmlrpages{}
\jmlryear{}
\makeatletter
\renewcommand*{\@titlefoot}{}
\makeatother

\title[The loss and recovery of scanner robustness in distillation]{Distilling Pathology Foundation Models for Mitotic Figure Classification: The Loss and Recovery of Scanner Robustness}

\midlauthor{\Name{Shenghan Gao} \Email{aaron.gao@wustl.edu}\\
 \Name{SeongJun Kwon} \Email{kwons@wustl.edu}\\
 \Name{Axel Ramirez} \Email{axel.r@wustl.edu}\\
 \addr Washington University in St.\ Louis, St.\ Louis, MO, USA}

\begin{document}

\maketitle

\begin{abstract}
Pathology foundation models classify mitotic figures robustly on unseen scanners but are too slow for every candidate
cell, so they are distilled into compact networks; whether the teacher's scanner robustness survives at a fixed
operating point is unknown. On the breast subset of MIDOG++, with
one of three scanners held out at a time, we distil four LoRA-adapted foundation models into ResNet-18 and ConvNeXt-T
students, decompose the student's loss into three layers (operating point, normalisation statistics, representation)
and repair each in order of cost, with pre-specified contrasts and a paired bootstrap over images and seeds.
Distillation transfers the teacher's ranking (AUROC) but not its robustness (F1). Most of the loss, however, needs no
retraining: for ResNet-18 and three of the four teachers, the operating point and BatchNorm statistics account for 51\%
to 75\% of the teacher's advantage, and for BatchNorm-free ConvNeXt-T the operating point alone accounts for 41\% to
46\%. One logit offset fitted to the teacher's predictions on unlabelled images (TGR) reaches the bound of a
labelled cutoff (within 0.005 F1 from five images on) and, unlike source-referenced recalibration and prior-shift
estimators, survives changes in mitosis prevalence. Retraining on unlabelled target images gives the largest gain,
mostly without the teacher, although for ConvNeXt-T the teacher adds 0.030 F1. Repaired layer by layer, a ConvNeXt-T
student with 23 to 41 times fewer parameters and 62 to 98 times the throughput recovers 86\% to 99\% of its teacher's
advantage and is non-inferior to Virchow2 at a margin of 0.02 F1; it stays 0.005 to 0.012 F1 below three LoRA
trainings of H-optimus-0, and non-inferiority to any one of them is not established.
\end{abstract}


\section{Introduction}
\label{sec:intro}

Pathology foundation models adapted with LoRA are among the strongest classifiers of mitotic figures on unseen domains
\citep{ammeling2026benchmarking}, but the four we use have 0.3 to 1.2
billion parameters and classify 80 to 342 cells per second on a workstation GPU, 68 to 287 times fewer than a
ResNet-18. Because a whole slide holds thousands of mitosis candidates, the natural route to deployment is to distil the
foundation model into a small task-specific network \citep{hinton2015distilling} that commits to one decision threshold.

Whether the teacher's scanner robustness survives this step is not known. Distilled pathology encoders keep most of
their teachers' average downstream performance, and one was reported to be more robust to scanner variation than its
teacher in embedding space \citep{filiot2025distilling,cai2026deployment,lenz2026multi}. These evaluations,
however, use threshold-free measures. Under scanner shift, the probabilities of foundation models drift even when their
AUC does not \citep{thiringer2026scanner,wen2026tempering}, so a threshold chosen on source scanners can fail on a new
scanner without any loss in ranking \citep{roschewitz2023automatic}. Embedding similarity and AUC hide this failure;
F1 at the deployed operating point exposes it. No study has measured, at that operating point, how much of a foundation
model's scanner robustness a small distilled student keeps, where the rest goes, and what it costs to get it back.

We do so for cell-level mitotic-figure classification on the breast subset of MIDOG++, holding out
one scanner at a time. We distil four LoRA-adapted foundation models into ResNet-18 and ConvNeXt-T students, describe the student's loss on the unseen scanner in three layers, the
\emph{operating point}, the \emph{normalisation statistics} and the \emph{representation}, and repair them in order of
cost (\figureref{fig:schematic}). We ask:
\textbf{RQ1} Does distillation carry the teacher's scanner robustness to the student?
\textbf{RQ2} Where does the student--teacher gap sit?
\textbf{RQ3} What does each layer need to be repaired, and what is the teacher needed for?
\textbf{RQ4} How close does the repaired student get to its teacher, and at what cost?
To separate the teacher's contribution from that of the unlabelled target images, we include controls without any
foundation model; to guard against selective reporting, every confirmatory contrast was written down, with its
expected direction, before it was tested.

\begin{figure}[t]
\floatconts
  {fig:schematic}
  {\caption{Study design. On an unseen scanner, the distilled student's loss is described in three layers, each repaired
  by a step that needs more than the previous one.}}
  {\includegraphics[width=0.9\linewidth]{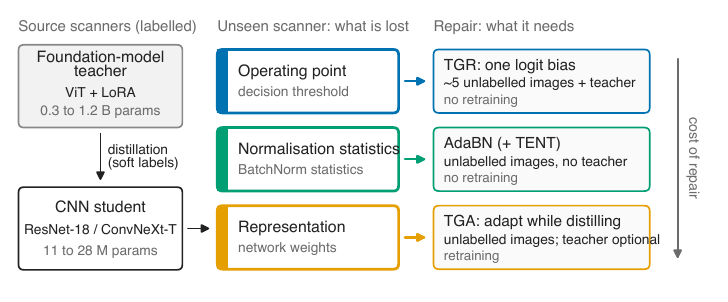}}
\end{figure}

We find: \textbf{(1) Distillation transfers ranking, not robustness.} No teacher gives the ResNet-18 student a
significant F1 gain on the unseen scanner, although its AUROC improves; the ConvNeXt-T student recovers 36\% to 63\% of
the teacher's advantage.
\textbf{(2) The gap is layered, and the layers depend on the student.} For ResNet-18, the operating point and
the BatchNorm statistics account for 51\% to 75\% of the teacher's advantage with three of the four teachers and are
repaired without retraining; for the BatchNorm-free ConvNeXt-T, the operating point alone accounts for 41\% to 46\%.
\textbf{(3) The teacher matters most at the operating point, where it supplies the target prevalence.} One logit
offset fitted to the teacher's predictions on unlabelled images (TGR) reaches the labelled-cutoff bound, to within
0.005 F1 from five images on, and, unlike source-referenced recalibration and prior-shift estimators, survives a change
in prevalence. Most of the retraining gain comes from the unlabelled images rather than from the teacher (75\% to 96\%
for ResNet-18, 57\% for ConvNeXt-T), though for ConvNeXt-T the teacher still adds 0.030 F1.
\textbf{(4) A repaired small student comes close to its teacher.} A ConvNeXt-T student with 23 to 41 times
fewer parameters and 62 to 98 times the throughput recovers 86\% to 99\% of its teacher's advantage and is non-inferior
to Virchow2 at a margin of 0.02 F1; against three LoRA trainings of H-optimus-0 it trails by 0.005 to 0.012 F1, and
non-inferiority to any one of them is not established. This is an analysis study: TGR is an intercept-only recalibration whose new element is its reference. The contribution
is the measured decomposition, the controls that attribute each repair to the teacher or to the target data, and the
cost-ordered guidance they give for deployment on new scanners.

\section{Related work}
\label{sec:related}

\paragraph{Mitoses and foundation models under domain shift.} Mitosis classifiers lose accuracy on unseen scanners and
tumour types \citep{aubreville2021quantifying,aubreville2023mitosis,aubreville2024domain}. At the cell level,
LoRA-adapted pathology foundation models are among the strongest out-of-domain classifiers
\citep{ammeling2026benchmarking}, yet their features still encode the scanner or centre
\citep{dejong2025current,komen2025towards}. On rescanned slides their AUC is largely stable while their probabilities
shift \citep{thiringer2026scanner}, so a fixed threshold changes its sensitivity \citep{wen2026tempering}; on MIDOG++,
F1 at a fixed threshold, calibration and count error likewise disagree under domain shift \citep{desouza2026metric}.

\paragraph{Distilling foundation models.} Pathology encoders distilled into ViT-B, ViT-S or convolutional students keep
most of their teachers' average downstream performance \citep{filiot2025distilling,cai2026deployment,lenz2026multi},
and H0-mini was reported to be more robust to scanner variation than its teacher in embedding space
\citep{filiot2025distilling}. In natural images, however, standard distillation transfers in-distribution accuracy more
reliably than out-of-distribution robustness \citep{zhou2023distilling,addepalli2024leveraging}, and
unlabelled target data help on some shifts but not consistently \citep{sagawa2022extending}. We therefore measure F1 at a
deployable operating point and separate the teacher's contribution from that of the target data.

\paragraph{Label-free recalibration and adaptation.} Existing label-free repairs of the operating point each rest on an
assumption that scanner shift can violate. UPA aligns target scores to source scores and so assumes preserved
prevalence \citep{roschewitz2023automatic}. Conversely, prior-shift methods estimate the target prevalence
\citep{saerens2002adjusting,lipton2018detecting,alexandari2020maximum} but assume unchanged class-conditional
distributions \citep{garg2020unified}, which acquisition shift breaks; prevalence shift itself degrades deployed
classifiers \citep{godau2023deployment}. One-parameter logit corrections from few unlabelled samples exist for other
tasks \citep{zhou2024batch,yang2026your}, and AdaBN and TENT adapt normalisation layers
\citep{li2018adaptive,schneider2020improving,wang2021tent}. TGR is an intercept-only correction too; its new element is
the reference, the teacher's predictions on the same target cells, which avoids both assumptions if the teacher is robust.

\section{Study design}
\label{sec:design}

\subsection{Data and evaluation protocol}

We use the human breast carcinoma subset of MIDOG++ \citep{aubreville2023comprehensive}: 150 cases, 50 from each of
three scanners (Hamamatsu XR, Hamamatsu S360, Aperio CS2), with 1{,}721 annotated mitotic figures and 2{,}714 hard
negatives (imposters), i.e.\ 4{,}435 cells. Each model classifies a patch centred on an annotation (mitotic figure vs.\
imposter); candidate detection is outside our scope. We hold out one scanner at a time and train on the official training images of the other two, except eight images per training scanner that
form a fixed validation set. Checkpoint and decision threshold (\emph{cutoff}) are always selected on this source
validation set by F1, so the held-out scanner never informs a training decision.

All 50 images of the held-out scanner are used for evaluation through cross-fitting. They are sorted by
identifier and split into alternating halves $A$ and $B$. Every method that uses unlabelled new-scanner data is fitted
on one half and applied to the other, then the halves are swapped, so each cell is scored by a model that never saw its
image in any form (except AdaBN + TGR; Appendix~\ref{app:impl}). The outcome is F1 on all cells of the held-out scanner, computed per run, averaged over the student
seeds and then over the three held-out scanners. New-scanner labels are used only for scoring and the
labelled-cutoff bound.

\subsection{Teachers and students}

\emph{Teachers.} Four pathology foundation models, H-optimus-0 \citep{hoptimus0}, Virchow2 \citep{zimmermann2024virchow2},
UNI \citep{chen2024uni} and UNI2-h \citep{uni2h}, are adapted with LoRA \citep{hu2022lora} as in a recent
mitotic-figure benchmark \citep{ammeling2026benchmarking} (rank 16, $\alpha=16$, dropout 0.1 on every attention and MLP
projection; frozen encoder), with a linear head on the class token concatenated with the mean patch token. Each teacher is
trained for 5 epochs on 224-pixel centre crops, once per held-out scanner and LoRA seed (three seeds).

\emph{Students.} To test whether the decomposition depends on BatchNorm, we use two ImageNet-initialised students:
ResNet-18 \citep{he2016deep} (11\,M parameters, BatchNorm) and ConvNeXt-T \citep{liu2022convnet} (28\,M parameters,
LayerNorm, no statistics to re-estimate), trained on 128-pixel crops for 60 epochs with 10 seeds per configuration. The
\emph{no-teacher student} is trained with expert labels only. The \emph{distilled (KD) student}
\citep{hinton2015distilling}, in contrast, never sees an expert label: it matches the teacher's temperature-softened
probability ($T=2$) and, through a linear projection, the teacher's feature vector (cosine loss, weight 1) on 16 cached
views of each source cell (8 dihedral transforms, each with and without H\&E stain jitter). Students are distilled from
the LoRA seed-0 teacher unless stated otherwise.

\subsection{Three layers and their repairs}
\label{sec:layers}

We describe what the student loses on a new scanner in three layers and repair them in order of cost
(\figureref{fig:schematic}; \tableref{tab:ladder} in Appendix~\ref{app:impl}).

\emph{Operating point.} Even when a student keeps its ranking (AUROC) on a new scanner, a cutoff chosen on source data
can be wrong there \citep{roschewitz2023automatic,thiringer2026scanner}. Our repair, \emph{teacher-referenced
recalibration} (TGR), keeps the student's weights and fits one logit offset so that the student agrees with the teacher
on the unlabelled cells $U$ of the fit half:
\begin{equation}
\hat b = \arg\min_b \; -\frac{1}{|U|}\sum_{x\in U}\Big[q(x)\log\sigma\big(z(x)+b\big) + \big(1-q(x)\big)\log\big(1-\sigma(z(x)+b)\big)\Big],
\label{eq:tgr}
\end{equation}
where $z(x)$ is the student logit and $q(x)$ the teacher probability; a cell of the other half is called positive if
$z(x)+\hat b \ge \operatorname{logit}(\tau_T)$, with $\tau_T$ the teacher's own source cutoff. Because the ranking is
kept, this shift carries the teacher's view of the target prevalence over to the student. TGR needs no retraining, only the
teacher's forward pass on a few unlabelled images (about 150 cells for five).

\emph{Normalisation statistics.} For ResNet-18, AdaBN \citep{li2018adaptive,schneider2020improving} re-estimates the
BatchNorm running statistics on the fit half without changing any learned weight. TENT \citep{wang2021tent} additionally
tunes the BatchNorm affine parameters by entropy minimisation (learning rate $10^{-3}$, one epoch; Appendix~\ref{app:tent}).

\emph{Representation.} Teacher-guided adaptation (TGA) retrains the KD student with the unlabelled cells of the fit
half added to the distillation set, using the teacher's answers and features on the same views. As a control for the
contribution of the teacher, \emph{no-teacher adaptation} trains on the same
unlabelled cells with the soft answers of an ensemble of the ten no-teacher ResNet-18 students after AdaBN on the same
half, and with expert labels on the source cells; no foundation model is involved, and the same pseudo-labeller serves
both students.

\subsection{Attribution}

With $F_\text{ctrl}$ the new-scanner F1 of the no-teacher student and $F_\text{teacher}$ that of the teacher at its
source cutoff, we express every arm by the \emph{share of the teacher's advantage recovered},
$(F_\text{arm}-F_\text{ctrl})/(F_\text{teacher}-F_\text{ctrl})$ (0\% = no-teacher student, 100\% = teacher), the
``performance gap recovered'' of \citet{burns2023weak}. Because AdaBN also moves the operating point, the credit of the
two post-hoc layers depends on their order. We therefore report Shapley values over \{threshold, statistics\}
\citep{zhang2023why}, i.e.\ each layer's gain averaged over the two orders. Retraining is credited with the gain of TGA beyond the two post-hoc layers, and the rest is the remaining gap.

\subsection{Statistical analysis}
\label{sec:stats}

Uncertainty comes from a paired bootstrap \citep{efron1993introduction} with $B=20{,}000$ replicates that resamples the
50 images of each held-out scanner (multinomial weights on images, shared by all arms) and the 10 student seeds, since
both are sources of variation \citep{varoquaux2022machine}; intervals are 95\% percentile intervals of the
fold-averaged statistic. Two-sided bootstrap $p$-values are adjusted with Holm's procedure \citep{holm1979simple} within
each family of contrasts.

Each family of confirmatory contrasts, together with its expected direction, was written down and its file hash
recorded before any of its contrasts was tested; each protocol states what already existed (Appendix~\ref{app:prereg});
results that contradict an expectation are reported as such, and analyses added afterwards are labelled descriptive. To
ask whether the student is as good as its teacher, we test non-inferiority with a margin of $\delta=0.02$ F1, the range
of the H-optimus-0 teacher's own new-scanner F1 over three LoRA seeds (0.801 to 0.821): the student is called
non-inferior if (student $-$ teacher) exceeds $-0.02$ at one-sided $\alpha=0.025$ after Holm correction within the
family. A failed test is reported as ``non-inferiority not established'', not as equivalence or
difference.

\section{Results}
\label{sec:results}

\begin{table}[t]
\floatconts{tab:main}
{\caption{F1 on the unseen scanner (mean over 10 student seeds, then over the three held-out scanners; 95\% interval from
a paired bootstrap over images and seeds). Each student uses its own teacher for TGR. --: not applicable (ConvNeXt-T has
no BatchNorm) or not run. Last row: (student $-$ no-teacher student) / (teacher $-$ no-teacher student).
$^\dagger$UNI's advantage over the no-teacher student is only 0.055 F1, so its share is unstable; its seed-0 LoRA
teacher chose an unusually low cutoff on the source scanners.}}
{%
\scriptsize
\setlength{\tabcolsep}{3pt}
\begin{tabular}{@{}lcccc@{\hspace{8pt}}cc@{}}
\toprule
 & \multicolumn{4}{c}{ResNet-18 student (11 M)} & \multicolumn{2}{c}{ConvNeXt-T student (28 M)} \\
\cmidrule(lr){2-5}\cmidrule(l){6-7}
Teacher & H-optimus-0 & Virchow2 & UNI2-h & UNI & H-optimus-0 & Virchow2 \\
\midrule
No-teacher student & \makecell{0.711\\{\scriptsize[.682, .735]}} & \makecell{0.711\\{\scriptsize[.682, .735]}} & \makecell{0.711\\{\scriptsize[.682, .735]}} & \makecell{0.711\\{\scriptsize[.682, .735]}} & \makecell{0.736\\{\scriptsize[.708, .759]}} & \makecell{0.736\\{\scriptsize[.708, .759]}} \\
Distillation (KD) & \makecell{0.725\\{\scriptsize[.696, .748]}} & \makecell{0.722\\{\scriptsize[.694, .744]}} & \makecell{0.704\\{\scriptsize[.673, .729]}} & \makecell{0.716\\{\scriptsize[.685, .741]}} & \makecell{0.765\\{\scriptsize[.740, .785]}} & \makecell{0.761\\{\scriptsize[.736, .781]}} \\
\quad + TGR & \makecell{0.762\\{\scriptsize[.739, .779]}} & \makecell{0.748\\{\scriptsize[.724, .767]}} & \makecell{0.744\\{\scriptsize[.720, .762]}} & \makecell{0.720\\{\scriptsize[.693, .741]}} & \makecell{0.801\\{\scriptsize[.781, .817]}} & \makecell{0.789\\{\scriptsize[.768, .806]}} \\
\quad + AdaBN + TGR & \makecell{0.777\\{\scriptsize[.752, .796]}} & \makecell{0.775\\{\scriptsize[.750, .793]}} & \makecell{0.768\\{\scriptsize[.745, .786]}} & \makecell{0.738\\{\scriptsize[.709, .759]}} & -- & -- \\
KD + TGA (retrained) & \makecell{0.788\\{\scriptsize[.768, .805]}} & \makecell{0.781\\{\scriptsize[.760, .798]}} & \makecell{0.771\\{\scriptsize[.750, .789]}} & \makecell{0.782\\{\scriptsize[.761, .799]}} & \makecell{0.805\\{\scriptsize[.784, .822]}} & \makecell{0.805\\{\scriptsize[.785, .821]}} \\
No-teacher adaptation & \makecell{0.769\\{\scriptsize[.744, .788]}} & \makecell{0.769\\{\scriptsize[.744, .788]}} & \makecell{0.769\\{\scriptsize[.744, .788]}} & \makecell{0.769\\{\scriptsize[.744, .788]}} & \makecell{0.775\\{\scriptsize[.749, .795]}} & \makecell{0.775\\{\scriptsize[.749, .795]}} \\
Teacher (seed 0) & \makecell{0.814\\{\scriptsize[.794, .830]}} & \makecell{0.805\\{\scriptsize[.785, .822]}} & \makecell{0.797\\{\scriptsize[.776, .813]}} & \makecell{0.767\\{\scriptsize[.743, .785]}} & \makecell{0.814\\{\scriptsize[.794, .830]}} & \makecell{0.805\\{\scriptsize[.785, .822]}} \\
\midrule
Share recovered by KD + TGA & \makecell{75\%\\{\scriptsize[64, 85]}} & \makecell{74\%\\{\scriptsize[62, 87]}} & \makecell{70\%\\{\scriptsize[54, 85]}} & \makecell{128\%$^\dagger$\\{\scriptsize[106, 174]}} & \makecell{88\%\\{\scriptsize[74, 104]}} & \makecell{99\%\\{\scriptsize[83, 121]}} \\
\bottomrule
\end{tabular}
}
\end{table}

\subsection{RQ1: distillation transfers ranking, not robustness}

The LoRA teachers reach a new-scanner F1 of 0.767 to 0.814 (\tableref{tab:main}; means over three LoRA seeds 0.778 to
0.812, comparable to unseen tumour types in \citealp{ammeling2026benchmarking}); the no-teacher ResNet-18 student reaches 0.711. Yet distilling any of the four teachers into ResNet-18 changes its F1 by only $-0.007$ to $+0.013$,
and every 95\% interval includes zero. ConvNeXt-T benefits more (+0.029, +0.014 to +0.045, with H-optimus-0; +0.025
with Virchow2) but recovers only 37\% and 36\% of the teacher's advantage (48\% and 63\% with two other
trainings of H-optimus-0). The shortfall is not a failure to rank cells: the AUROC of the distilled ResNet-18
students (0.879 to 0.894) exceeds that of the no-teacher student (0.872) and follows the teachers' order, so the loss in
F1 lies elsewhere.

\subsection{RQ2: where the gap sits depends on the student}

\begin{figure}[t]
\floatconts
  {fig:gap}
  {\caption{Where the gap sits. Each bar splits the teacher's advantage over the no-teacher student (0\% to 100\%) into
  distillation, operating point (TGR), normalisation statistics (AdaBN; Shapley credit over the two post-hoc layers),
  retraining (TGA beyond the post-hoc layers) and the remaining gap; triangles: retrained student with 95\% interval.
  UNI's advantage is too small for a stable share (\tableref{tab:main}).}}
  {\includegraphics[width=\linewidth]{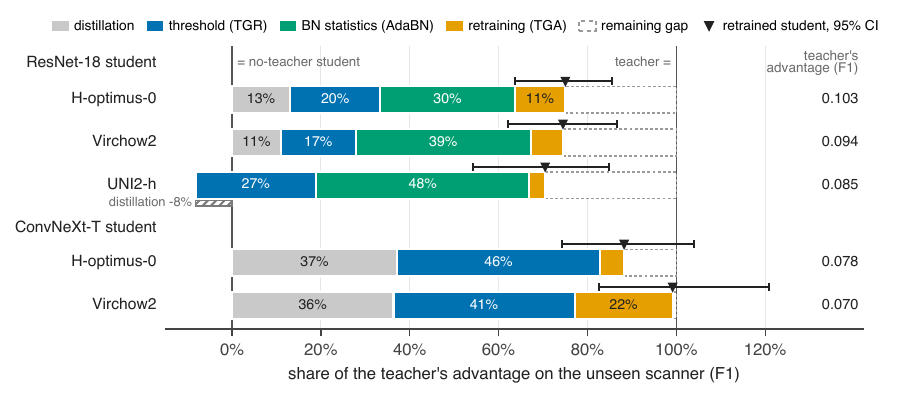}}
\end{figure}

For the ResNet-18 student, the two post-hoc layers recover most of the teacher's advantage without retraining
(\figureref{fig:gap}): the operating point accounts for 17\% to 27\% and the BatchNorm statistics for 30\% to 48\%,
together 51\% to 75\% for the three teachers with a reliable operating point. Once both are repaired, retraining adds
only 4\% to 11\%, and 25\% to 30\% remains. ConvNeXt-T, by contrast, has no statistics to re-estimate; there the
operating point alone accounts for 46\% (H-optimus-0) and 41\% (Virchow2), and retraining adds 5\% and 22\%. Retraining is not disguised AdaBN: putting source
statistics back into the retrained ResNet-18 student costs only 0.011 (S360) and 0.007 F1 (XR)
(Appendix~\ref{app:ladder}).

\subsection{RQ3a: the operating point needs the teacher, but only a few images of it}
\label{sec:res-threshold}

\begin{figure}[t]
\floatconts
  {fig:threshold}
  {\caption{The operating point. (a, b) F1 against the number of new-scanner images per half, used unlabelled (TGR,
  AdaBN) or labelled (labelled cutoff); means over folds, seeds and 10 image draws, bands: range over draws. (c, d) F1
  after resampling new-scanner cells to a given mitosis prevalence; grey: source prevalence range.}}
  {\includegraphics[width=0.95\linewidth]{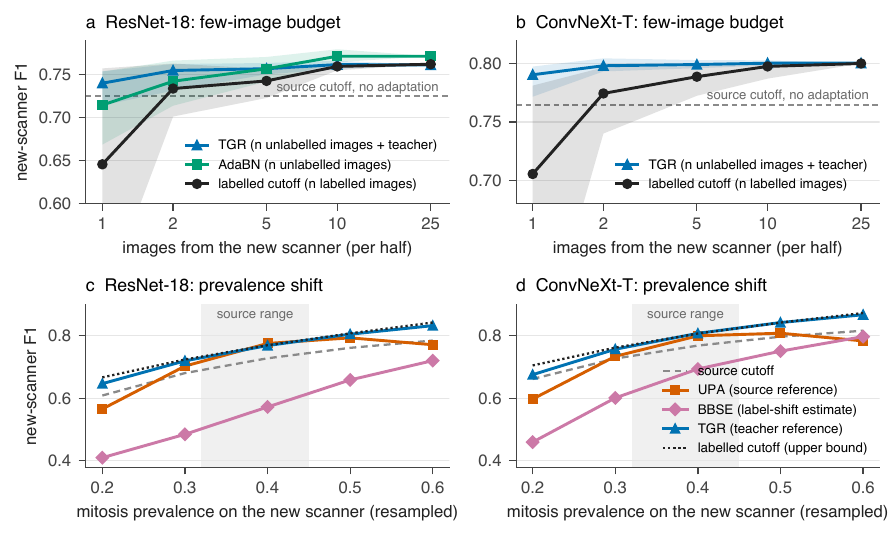}}
\end{figure}

One teacher-referenced offset closes the threshold gap from few images: TGR brings both KD students to the bound of a cutoff chosen with the new scanner's labels (ResNet-18: 0.762 vs.\
0.762; ConvNeXt-T: 0.801 vs.\ 0.800), +0.037 (+0.022 to +0.053) and +0.036 (+0.022 to +0.052) over the source cutoff
(Holm $p<0.001$), and it comes within 0.005 of that bound from five unlabelled images per half on
(\figureref{fig:threshold}a,b). With a single image it beats AdaBN, whose statistics cannot be estimated from so few
cells (+0.025, +0.010 to +0.041, Holm $p=0.004$), and at every budget it matches or beats a cutoff chosen from the labels
of the same images (one image: 0.740 vs.\ 0.646; Appendix~\ref{app:few}).

What the teacher contributes here is the target prevalence. At the natural prevalence, close to that of the source
scanners, keeping the source positive rate (what UPA reduces to at one operating point) is only 0.002 below TGR for ResNet-18 ($-0.008$ to $+0.012$, contrary to our expectation) and 0.019
below for ConvNeXt-T (+0.009 to +0.031, Holm $p<0.001$). Once the prevalence changes, however, source-referenced
methods break while TGR stays near the labelled bound (\figureref{fig:threshold}c,d): at 0.2 or 0.6 it beats the source positive rate by 0.061 to 0.083 (four pre-specified contrasts, Holm $p<0.001$) and the same
offset referenced to the source mean probability by 0.025 to 0.126. Three of the four other teacher-referenced recalibrations stay
within 0.015 of TGR (Appendix~\ref{app:baselines}), so the reference matters more than the functional form. Prior-shift
estimators, conversely, fail even at the natural prevalence (EM, EM with bias-corrected calibration, BBSE: 0.52 to 0.55
for ResNet-18): on S360 the student calls 21\% of cells positive (true prevalence 35\%) and they conclude 4\% to 9\%,
reading the scanner-induced score shift as a prevalence change.

\subsection{RQ3b: retraining gains come mostly from the unlabelled images}

Retraining with the teacher on the unlabelled half (TGA) is the largest step: +0.060 to +0.067 over the KD
ResNet-18 student for every teacher (Holm $p<0.001$ for the three pre-specified ones), and +0.040 (H-optimus-0) and +0.044 (Virchow2; both $p<0.001$)
for ConvNeXt-T. Most of this gain, however, does not need the teacher. For ResNet-18, the target images alone reach 75\% to 96\% of TGA's gain over the no-teacher student (0.769),
and the teacher adds +0.002 to +0.019 (not significant after Holm correction for the three pre-specified teachers;
+0.019, +0.006 to +0.034, for H-optimus-0, analysed before the protocols were written). AdaBN, TENT and TGR together
reach 0.784 without retraining, 0.005 ($-0.005$ to $+0.016$) below TGA. For ConvNeXt-T, the target images alone still
reach most of TGA's gain, 57\% (34\% to 76\%; 0.775), but the teacher adds a significant +0.030 with both teachers (+0.016 to +0.045 and +0.017 to +0.046, Holm
$p<0.001$). Here the adapted no-teacher student even stays below KD + TGR without retraining ($-0.025$, $-0.014$;
descriptive), perhaps limited by its ResNet-18 pseudo-labels.

\subsection{RQ4: a student 23--41 times smaller close to its teacher}

\begin{figure}[t]
\floatconts
  {fig:close}
  {\caption{How close and how cheap. Left: student minus the teacher training it was distilled from (LoRA seed 0 unless
  stated), F1 with 95\% intervals; filled: pre-specified (margin $-0.02$, dashed; brown: mean of three trainings);
  open: descriptive. Right: throughput.}}
  {\includegraphics[width=0.95\linewidth]{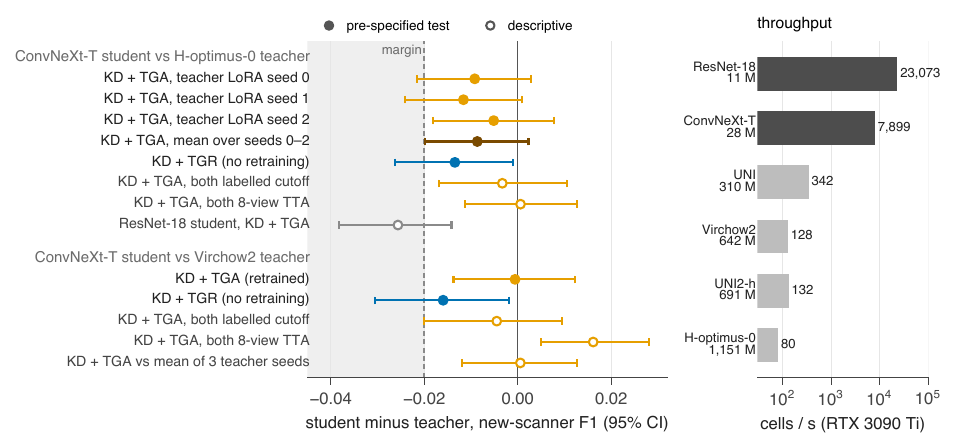}}
\end{figure}

Repaired in all three layers, the ConvNeXt-T student recovers 88\% (74\% to 104\%)
of H-optimus-0's advantage and 99\% (83\% to 121\%) of Virchow2's. It is non-inferior to Virchow2 at the pre-specified
margin ($-0.001$, $-0.014$ to $+0.012$; Holm $p=0.004$). Against H-optimus-0 it is 0.009 lower ($-0.022$ to $+0.003$),
so non-inferiority is not established ($p=0.085$; \figureref{fig:close}), nor for students of two other LoRA
trainings of this teacher (pre-specified, Appendix~\ref{app:fc}): they recover 86\% and 92\% of their own teacher's
advantage and stay 0.012 ($-0.024$ to $+0.001$) and 0.005 ($-0.018$ to $+0.008$) below it (Holm $p=0.090$ and
$0.027$; we expected non-inferiority for seed 2). On average over the three trainings (pre-specified) the gap is 0.009 (lower limit
$-0.0198$, just inside the margin), half the spread of the teacher's own F1 across them (0.020). With TGR alone, the KD
students stay below both seed-0 teachers ($-0.013$, $-0.016$), and the retrained ResNet-18 student remains 0.026
($-0.038$ to $-0.014$) below H-optimus-0. Neither the operating point nor test-time augmentation explains this: with labelled cutoffs on both sides
the differences are $-0.003$ (H-optimus-0) and $-0.004$ (Virchow2), and eight-view augmentation does not help the
teachers (0.814 to 0.812; 0.805 to 0.797). With 23 to 41 times fewer parameters, the
ConvNeXt-T student classifies 7{,}899 cells per second, 62 and 98 times as many as Virchow2 and H-optimus-0 (8 to 12
times if only the student uses eight-view augmentation), and the teacher is needed once per new scanner.

\section{Discussion}
\label{sec:discussion}

\paragraph{What the decomposition shows.} Distilled students can look robust by embedding or ranking measures
\citep{filiot2025distilling} yet fail at a fixed operating point: they keep the teacher's ranking but not the location
of its scores on a new scanner. The operating point fails first and, for a BatchNorm network, the
normalisation statistics second; both are restored from unlabelled images without retraining. Only the
representation gap needs retraining, and for ResNet-18 about a quarter of the teacher's advantage remains that none of
our steps closes. The teacher is needed mainly at the operating point, where its predictions carry the target
prevalence that no source-referenced method can know; in retraining, its soft labels help ConvNeXt-T more than
ResNet-18 (+0.030 vs.\ +0.002 to +0.019 F1).

\paragraph{Practical guidance.} With one or two unlabelled images, recalibrate against the teacher
(TGR); with one image this beats re-estimating BatchNorm statistics, and at every budget it matches or beats a cutoff
chosen from as many labelled images. With five or more images and a BatchNorm student, re-estimate the statistics, then
recalibrate; a BatchNorm-free student should be recalibrated first. If compute allows, retrain on the unlabelled images.
If the target prevalence may differ from the source, avoid source-referenced recalibration, and do not trust
prior-shift estimates under acquisition shift.

\paragraph{Relation to previous work.} At a deployed operating point, a distilled model is less robust than its
teacher, as in natural images \citep{addepalli2024leveraging}. Teacher-referenced recalibration drops UPA's assumption of preserved prevalence
\citep{roschewitz2023automatic}, and prior-shift estimators fail as acquisition shift breaks their assumptions
\citep{garg2020unified}.

\paragraph{Limitations.} All results come from one dataset, one organ and three scanners; their consistency over four
teachers, two students and three held-out scanners supports them within this setting, not beyond it. The prevalence
shift was simulated by resampling, which changes the class mix but not cell appearance. We classify given candidates and
do not detect them. The ResNet-18 pseudo-labels of the no-teacher control may understate
what target images alone can do for ConvNeXt-T. TGR inherits its teacher's operating point and fails when that is poor (UNI,
Appendix~\ref{app:ladder}). Except for ConvNeXt-T with H-optimus-0, each student was distilled from one LoRA training of
its teacher.

\paragraph{Conclusion.} The scanner robustness lost in distillation sits in three layers, the operating point, the
normalisation statistics and the representation, each repaired at a different cost. Repaired layer by layer, mostly
from a few unlabelled images, a student 23 to 41 times smaller recovers 86\% to 99\% of its teacher's advantage on an
unseen scanner.

\bibliography{refs}

\begin{thebibliography}{40}
\providecommand{\natexlab}[1]{#1}
\providecommand{\url}[1]{\texttt{#1}}
\expandafter\ifx\csname urlstyle\endcsname\relax
  \providecommand{\doi}[1]{doi: #1}\else
  \providecommand{\doi}{doi: \begingroup \urlstyle{rm}\Url}\fi

\bibitem[Addepalli et~al.(2024)Addepalli, Asokan, Sharma, and
  Babu]{addepalli2024leveraging}
Sravanti Addepalli, Ashish~Ramayee Asokan, Lakshay Sharma, and R.~Venkatesh
  Babu.
\newblock {Leveraging Vision-Language Models for Improving Domain
  Generalization in Image Classification}.
\newblock In \emph{IEEE/CVF Conference on Computer Vision and Pattern
  Recognition (CVPR)}, 2024.

\bibitem[Alexandari et~al.(2020)Alexandari, Kundaje, and
  Shrikumar]{alexandari2020maximum}
Amr Alexandari, Anshul Kundaje, and Avanti Shrikumar.
\newblock {Maximum Likelihood with Bias-Corrected Calibration is Hard-To-Beat
  at Label Shift Adaptation}.
\newblock In \emph{International Conference on Machine Learning (ICML)}, 2020.

\bibitem[Ammeling et~al.(2026)Ammeling, Ganz, Rosbach, Lausser, Bertram,
  Breininger, and Aubreville]{ammeling2026benchmarking}
Jonas Ammeling, Jonathan Ganz, Emely Rosbach, Ludwig Lausser, Christof~A.
  Bertram, Katharina Breininger, and Marc Aubreville.
\newblock {Benchmarking Foundation Models for Mitotic Figure Classification}.
\newblock \emph{Machine Learning for Biomedical Imaging}, 2026.
\newblock \doi{10.59275/j.melba.2026-a3eb}.

\bibitem[Aubreville et~al.(2021)Aubreville, Bertram, Veta, Klopfleisch,
  Stathonikos, Breininger, {ter Hoeve}, Ciompi, and
  Maier]{aubreville2021quantifying}
Marc Aubreville, Christof Bertram, Mitko Veta, Robert Klopfleisch, Nikolas
  Stathonikos, Katharina Breininger, Natalie {ter Hoeve}, Francesco Ciompi, and
  Andreas Maier.
\newblock {Quantifying the Scanner-Induced Domain Gap in Mitosis Detection}.
\newblock arXiv:2103.16515, 2021.

\bibitem[Aubreville et~al.(2023{\natexlab{a}})Aubreville, Stathonikos, Bertram,
  Klopfleisch, {ter Hoeve}, Ciompi, Wilm, Marzahl, Donovan, Maier,
  et~al.]{aubreville2023mitosis}
Marc Aubreville, Nikolas Stathonikos, Christof~A. Bertram, Robert Klopfleisch,
  Natalie {ter Hoeve}, Francesco Ciompi, Frauke Wilm, Christian Marzahl,
  Taryn~A. Donovan, Andreas Maier, et~al.
\newblock {Mitosis domain generalization in histopathology images -- The MIDOG
  challenge}.
\newblock \emph{Medical Image Analysis}, 84:\penalty0 102699,
  2023{\natexlab{a}}.
\newblock \doi{10.1016/j.media.2022.102699}.

\bibitem[Aubreville et~al.(2023{\natexlab{b}})Aubreville, Wilm, Stathonikos,
  Breininger, Donovan, Jabari, Veta, Ganz, Ammeling, van Diest,
  et~al.]{aubreville2023comprehensive}
Marc Aubreville, Frauke Wilm, Nikolas Stathonikos, Katharina Breininger,
  Taryn~A. Donovan, Samir Jabari, Mitko Veta, Jonathan Ganz, Jonas Ammeling,
  Paul~J. van Diest, et~al.
\newblock {A comprehensive multi-domain dataset for mitotic figure detection}.
\newblock \emph{Scientific Data}, 10:\penalty0 484, 2023{\natexlab{b}}.
\newblock \doi{10.1038/s41597-023-02327-4}.

\bibitem[Aubreville et~al.(2024)Aubreville, Stathonikos, Donovan, Klopfleisch,
  Ammeling, Ganz, Wilm, Veta, Jabari, Eckstein, et~al.]{aubreville2024domain}
Marc Aubreville, Nikolas Stathonikos, Taryn~A. Donovan, Robert Klopfleisch,
  Jonas Ammeling, Jonathan Ganz, Frauke Wilm, Mitko Veta, Samir Jabari, Markus
  Eckstein, et~al.
\newblock {Domain generalization across tumor types, laboratories, and species
  -- insights from the 2022 edition of the Mitosis Domain Generalization
  Challenge}.
\newblock \emph{Medical Image Analysis}, 94:\penalty0 103155, 2024.
\newblock \doi{10.1016/j.media.2024.103155}.

\bibitem[Burns et~al.(2024)Burns, Izmailov, Kirchner, Baker, Gao,
  Aschenbrenner, Chen, Ecoffet, Joglekar, Leike, et~al.]{burns2023weak}
Collin Burns, Pavel Izmailov, Jan~Hendrik Kirchner, Bowen Baker, Leo Gao,
  Leopold Aschenbrenner, Yining Chen, Adrien Ecoffet, Manas Joglekar, Jan
  Leike, et~al.
\newblock {Weak-to-Strong Generalization: Eliciting Strong Capabilities With
  Weak Supervision}.
\newblock In \emph{International Conference on Machine Learning (ICML)}, 2024.

\bibitem[Cai et~al.(2026)Cai, Jin, Zhang, Ma, Zhou, Xu, Guo, Wang, Zhang,
  Liang, et~al.]{cai2026deployment}
Yu~Cai, Cheng Jin, Zhengyu Zhang, Jiabo Ma, Fengtao Zhou, Yingxue Xu, Zhengrui
  Guo, Yihui Wang, Zhengyu Zhang, Ling Liang, et~al.
\newblock {A Deployment-Friendly Foundational Framework for Efficient
  Computational Pathology}.
\newblock arXiv:2602.14010, 2026.

\bibitem[Chen et~al.(2024)Chen, Ding, Lu, Williamson, Jaume, Song, Chen, Zhang,
  Shao, Shaban, et~al.]{chen2024uni}
Richard~J. Chen, Tong Ding, Ming~Y. Lu, Drew F.~K. Williamson, Guillaume Jaume,
  Andrew~H. Song, Bowen Chen, Andrew Zhang, Daniel Shao, Muhammad Shaban,
  et~al.
\newblock {Towards a general-purpose foundation model for computational
  pathology}.
\newblock \emph{Nature Medicine}, 30:\penalty0 850--862, 2024.
\newblock \doi{10.1038/s41591-024-02857-3}.

\bibitem[{de Jong} et~al.(2025){de Jong}, Marcus, and
  Teuwen]{dejong2025current}
Edwin~D. {de Jong}, Eric Marcus, and Jonas Teuwen.
\newblock {Current Pathology Foundation Models are unrobust to Medical Center
  Differences}.
\newblock arXiv:2501.18055, 2025.

\bibitem[de~Souza(2026)]{desouza2026metric}
Cleverson de~Souza.
\newblock {Metric discordance under MIDOG++ domain shift: a
  leave-one-domain-out audit of F1, emitted-detection confidence calibration,
  and ROI count error}.
\newblock \emph{Journal of Pathology Informatics}, 22:\penalty0 100705, 2026.
\newblock \doi{10.1016/j.jpi.2026.100705}.

\bibitem[Efron and Tibshirani(1993)]{efron1993introduction}
Bradley Efron and Robert~J. Tibshirani.
\newblock \emph{{An Introduction to the Bootstrap}}.
\newblock Chapman \& Hall/CRC, 1993.

\bibitem[Filiot et~al.(2025)Filiot, Dop, Tchita, Riou, Dubois, Peeters, Valter,
  Scalbert, Saillard, Robin, et~al.]{filiot2025distilling}
Alexandre Filiot, Nicolas Dop, Oussama Tchita, Auriane Riou, Rémy Dubois,
  Thomas Peeters, Daria Valter, Marin Scalbert, Charlie Saillard, Geneviève
  Robin, et~al.
\newblock {Distilling foundation models for robust and efficient models in
  digital pathology}.
\newblock In \emph{Medical Image Computing and Computer Assisted Intervention
  (MICCAI)}, 2025.
\newblock arXiv:2501.16239.

\bibitem[Garg et~al.(2020)Garg, Wu, Balakrishnan, and Lipton]{garg2020unified}
Saurabh Garg, Yifan Wu, Sivaraman Balakrishnan, and Zachary~C. Lipton.
\newblock {A Unified View of Label Shift Estimation}.
\newblock In \emph{Advances in Neural Information Processing Systems
  (NeurIPS)}, 2020.

\bibitem[Godau et~al.(2023)Godau, Kalinowski, Christodoulou, Reinke, Tizabi,
  Ferrer, Jäger, and Maier-Hein]{godau2023deployment}
Patrick Godau, Piotr Kalinowski, Evangelia Christodoulou, Annika Reinke, Minu
  Tizabi, Luciana Ferrer, Paul Jäger, and Lena Maier-Hein.
\newblock {Deployment of Image Analysis Algorithms under Prevalence Shifts}.
\newblock In \emph{Medical Image Computing and Computer Assisted Intervention
  (MICCAI)}, Lecture Notes in Computer Science, pages 389--399, 2023.
\newblock \doi{10.1007/978-3-031-43898-1_38}.

\bibitem[He et~al.(2016)He, Zhang, Ren, and Sun]{he2016deep}
Kaiming He, Xiangyu Zhang, Shaoqing Ren, and Jian Sun.
\newblock {Deep Residual Learning for Image Recognition}.
\newblock In \emph{IEEE Conference on Computer Vision and Pattern Recognition
  (CVPR)}, 2016.

\bibitem[Hinton et~al.(2015)Hinton, Vinyals, and Dean]{hinton2015distilling}
Geoffrey Hinton, Oriol Vinyals, and Jeff Dean.
\newblock {Distilling the Knowledge in a Neural Network}.
\newblock arXiv:1503.02531, 2015.

\bibitem[Holm(1979)]{holm1979simple}
Sture Holm.
\newblock {A simple sequentially rejective multiple test procedure}.
\newblock \emph{Scandinavian Journal of Statistics}, 6\penalty0 (2):\penalty0
  65--70, 1979.

\bibitem[Hu et~al.(2022)Hu, Shen, Wallis, Allen-Zhu, Li, Wang, Wang, and
  Chen]{hu2022lora}
Edward~J. Hu, Yelong Shen, Phillip Wallis, Zeyuan Allen-Zhu, Yuanzhi Li, Shean
  Wang, Lu~Wang, and Weizhu Chen.
\newblock {LoRA: Low-Rank Adaptation of Large Language Models}.
\newblock In \emph{International Conference on Learning Representations
  (ICLR)}, 2022.

\bibitem[Kömen et~al.(2026)Kömen, {de Jong}, Hense, Marienwald, Dippel,
  Naumann, Marcus, Ruff, Alber, Teuwen, et~al.]{komen2025towards}
Jonah Kömen, Edwin~D. {de Jong}, Julius Hense, Hannah Marienwald, Jonas
  Dippel, Philip Naumann, Eric Marcus, Lukas Ruff, Maximilian Alber, Jonas
  Teuwen, et~al.
\newblock {Towards robust foundation models for digital pathology}.
\newblock \emph{Nature Communications}, 17:\penalty0 5218, 2026.
\newblock \doi{10.1038/s41467-026-73923-2}.

\bibitem[Lenz et~al.(2026)Lenz, Heide, Gustav, Reitsam, and
  Kather]{lenz2026multi}
Tim Lenz, Maurice Heide, Marco Gustav, Nic~G. Reitsam, and Jakob~Nikolas
  Kather.
\newblock {Multi-Teacher Contrastive Distillation for Edge-Efficient Pathology
  Foundation Models}.
\newblock arXiv:2607.05533, 2026.

\bibitem[Li et~al.(2018)Li, Wang, Shi, Hou, and Liu]{li2018adaptive}
Yanghao Li, Naiyan Wang, Jianping Shi, Xiaodi Hou, and Jiaying Liu.
\newblock {Adaptive batch normalization for practical domain adaptation}.
\newblock \emph{Pattern Recognition}, 80:\penalty0 109--117, 2018.
\newblock \doi{10.1016/j.patcog.2018.03.005}.

\bibitem[Lipton et~al.(2018)Lipton, Wang, and Smola]{lipton2018detecting}
Zachary~C. Lipton, Yu-Xiang Wang, and Alex Smola.
\newblock {Detecting and Correcting for Label Shift with Black Box Predictors}.
\newblock In \emph{International Conference on Machine Learning (ICML)}, 2018.

\bibitem[Liu et~al.(2022)Liu, Mao, Wu, Feichtenhofer, Darrell, and
  Xie]{liu2022convnet}
Zhuang Liu, Hanzi Mao, Chao-Yuan Wu, Christoph Feichtenhofer, Trevor Darrell,
  and Saining Xie.
\newblock {A ConvNet for the 2020s}.
\newblock In \emph{IEEE/CVF Conference on Computer Vision and Pattern
  Recognition (CVPR)}, 2022.

\bibitem[MahmoodLab(2025)]{uni2h}
MahmoodLab.
\newblock {UNI2-h}.
\newblock \url{https://huggingface.co/MahmoodLab/UNI2-h}, 2025.

\bibitem[Roschewitz et~al.(2023)Roschewitz, Khara, Yearsley, Sharma,
  et~al.]{roschewitz2023automatic}
M{\'e}lanie Roschewitz, Galvin Khara, Joe Yearsley, Nisha Sharma, et~al.
\newblock {Automatic correction of performance drift under acquisition shift in
  medical image classification}.
\newblock \emph{Nature Communications}, 14:\penalty0 6608, 2023.
\newblock \doi{10.1038/s41467-023-42396-y}.

\bibitem[Saerens et~al.(2002)Saerens, Latinne, and
  Decaestecker]{saerens2002adjusting}
Marco Saerens, Patrice Latinne, and Christine Decaestecker.
\newblock {Adjusting the outputs of a classifier to new a priori probabilities:
  a simple procedure}.
\newblock \emph{Neural Computation}, 14\penalty0 (1):\penalty0 21--41, 2002.
\newblock \doi{10.1162/089976602753284446}.

\bibitem[Sagawa et~al.(2022)Sagawa, Koh, Lee, Gao, Xie, Shen, Kumar, Hu,
  Yasunaga, Marklund, et~al.]{sagawa2022extending}
Shiori Sagawa, Pang~Wei Koh, Tony Lee, Irena Gao, Sang~Michael Xie, Kendrick
  Shen, Ananya Kumar, Weihua Hu, Michihiro Yasunaga, Henrik Marklund, et~al.
\newblock {Extending the WILDS Benchmark for Unsupervised Adaptation}.
\newblock In \emph{International Conference on Learning Representations
  (ICLR)}, 2022.

\bibitem[Saillard et~al.(2024)Saillard, Jenatton, Llinares-L{\'o}pez, Mariet,
  Cahan{\'e}, Durand, and Vert]{hoptimus0}
Charlie Saillard, Rodolphe Jenatton, Felipe Llinares-L{\'o}pez, Zelda Mariet,
  David Cahan{\'e}, Eric Durand, and Jean-Philippe Vert.
\newblock {H-optimus-0}.
\newblock
  \url{https://github.com/bioptimus/releases/tree/main/models/h-optimus/v0},
  2024.

\bibitem[Schneider et~al.(2020)Schneider, Rusak, Eck, Bringmann, Brendel, and
  Bethge]{schneider2020improving}
Steffen Schneider, Evgenia Rusak, Luisa Eck, Oliver Bringmann, Wieland Brendel,
  and Matthias Bethge.
\newblock {Improving robustness against common corruptions by covariate shift
  adaptation}.
\newblock In \emph{Advances in Neural Information Processing Systems
  (NeurIPS)}, 2020.

\bibitem[Thiringer et~al.(2026)Thiringer, Gustafsson, Eriksson, and
  Rantalainen]{thiringer2026scanner}
Erik Thiringer, Fredrik~K. Gustafsson, Kajsa~Ledesma Eriksson, and Mattias
  Rantalainen.
\newblock {Scanner-Induced Domain Shifts Undermine the Robustness of Pathology
  Foundation Models}.
\newblock arXiv:2601.04163, 2026.

\bibitem[Varoquaux and Cheplygina(2022)]{varoquaux2022machine}
Ga{\"e}l Varoquaux and Veronika Cheplygina.
\newblock {Machine learning for medical imaging: methodological failures and
  recommendations for the future}.
\newblock \emph{npj Digital Medicine}, 5:\penalty0 48, 2022.
\newblock \doi{10.1038/s41746-022-00592-y}.

\bibitem[Wang et~al.(2021)Wang, Shelhamer, Liu, Olshausen, and
  Darrell]{wang2021tent}
Dequan Wang, Evan Shelhamer, Shaoteng Liu, Bruno Olshausen, and Trevor Darrell.
\newblock {Tent: Fully Test-time Adaptation by Entropy Minimization}.
\newblock In \emph{International Conference on Learning Representations
  (ICLR)}, 2021.

\bibitem[Wen et~al.(2026)Wen, Rosen, Ramesh, Koo, Kochanny, Jain-Liu, Dyer,
  Pearson, and Dolezal]{wen2026tempering}
Frank~T Wen, Kyler Rosen, Siddhi Ramesh, Michael Koo, Sara Kochanny, Sasha
  Jain-Liu, Emma Dyer, Alexander~T Pearson, and James~M Dolezal.
\newblock {Tempering and few-shot calibration of pathology foundation models
  improve digital biomarker reproducibility across slide scanners}.
\newblock arXiv:2611.31536, 2026.

\bibitem[Yang et~al.(2026)Yang, Goenawan, Wang, Qin, Xu, Yang, Fang, Sun, Lim,
  and Zhu]{yang2026your}
Muli Yang, Gabriel~James Goenawan, Henan Wang, Huaiyuan Qin, Chenghao Xu,
  Yanhua Yang, Fen Fang, Ying Sun, Joo-Hwee Lim, and Hongyuan Zhu.
\newblock {Your AI-Generated Image Detector Can Secretly Achieve SOTA Accuracy,
  If Calibrated}.
\newblock In \emph{AAAI Conference on Artificial Intelligence}, pages
  11622--11630, 2026.

\bibitem[Zhang et~al.(2023)Zhang, Singh, Ghassemi, and Joshi]{zhang2023why}
Haoran Zhang, Harvineet Singh, Marzyeh Ghassemi, and Shalmali Joshi.
\newblock {"Why did the Model Fail?": Attributing Model Performance Changes to
  Distribution Shifts}.
\newblock In \emph{International Conference on Machine Learning (ICML)}, 2023.

\bibitem[Zhou et~al.(2023)Zhou, Wang, Wang, and Wang]{zhou2023distilling}
Andy Zhou, Jindong Wang, Yu-Xiong Wang, and Haohan Wang.
\newblock {Distilling Out-of-Distribution Robustness from Vision-Language
  Foundation Models}.
\newblock In \emph{Advances in Neural Information Processing Systems
  (NeurIPS)}, 2023.

\bibitem[Zhou et~al.(2024)Zhou, Wan, Proleev, Mincu, Chen, Heller, and
  Roy]{zhou2024batch}
Han Zhou, Xingchen Wan, Lev Proleev, Diana Mincu, Jilin Chen, Katherine Heller,
  and Subhrajit Roy.
\newblock {Batch Calibration: Rethinking Calibration for In-Context Learning
  and Prompt Engineering}.
\newblock In \emph{International Conference on Learning Representations
  (ICLR)}, 2024.

\bibitem[Zimmermann et~al.(2024)Zimmermann, Vorontsov, Viret, Casson,
  Zelechowski, Shaikovski, Tenenholtz, Hall, Klimstra, Yousfi,
  et~al.]{zimmermann2024virchow2}
Eric Zimmermann, Eugene Vorontsov, Julian Viret, Adam Casson, Michal
  Zelechowski, George Shaikovski, Neil Tenenholtz, James Hall, David Klimstra,
  Razik Yousfi, et~al.
\newblock {Virchow2: Scaling Self-Supervised Mixed Magnification Models in
  Pathology}.
\newblock arXiv:2408.00738, 2024.

\end{thebibliography}

\appendix
\section{Pre-registration and analysis log}
\label{app:prereg}

Every confirmatory family was written into a protocol file before any of its contrasts was tested, i.e.\ before any
interval or $p$-value it concerns was computed; the SHA-256 hash of the file and the time were recorded at that moment
(\tableref{tab:app-protocols}). Each protocol states what already existed: for R, T, P, N1 and C some or all of the
models had been trained, and for the ResNet-18 rows of R1 and for N1 the point estimates were known (marked $^\ast$ in
\tableref{tab:app-tests}).
Later edits are listed as amendments with their own hashes. \tableref{tab:app-tests} lists every pre-specified contrast
with its expectation and outcome. Its $p$-values are Holm-adjusted within the family and two-sided, except that they
are one-sided for the non-inferiority contrasts (N1a, N1b, N2d, N2e, F1, F2), unadjusted for the single contrasts E1.7a
(amendment 1 of E1.7) and N2c, and unadjusted for the key secondary estimate of F, which is outside the Holm family
and includes the already-known seed-0 term. An expectation counts as met if the difference is significant in the
expected direction (for non-inferiority: if it is established), unless the expectation says otherwise. Five expectations were not met: four had the expected sign but were not significant (TGR
against the source positive rate and against AdaBN for ResNet-18 at the natural prevalence, TENT against KD + TGA, and
TGR against the source cutoff for ConvNeXt-T at $\pi=0.2$), and non-inferiority to the seed-2 training of H-optimus-0
(F2) was not established; two were met only in part (TENT against AdaBN) and one
result was better than expected (N2d).
The ResNet-18 arms with the H-optimus-0 teacher (no-teacher student, KD, KD + TGA, no-teacher adaptation) were run first
and analysed before these protocols; the expectations of the later families were derived from them.

Analyses not covered by a protocol are descriptive. Besides those labelled in the text, they are: the attribution and
the share of the teacher's advantage (\figureref{fig:gap}, including the recovery shares in RQ4), the additional
recalibration baselines (UPA, one bias to the source mean, EM with bias-corrected calibration, BBSE, bias + temperature,
quantile mapping to the teacher and the teacher positive rate), AdaBN + TENT + TGR, the comparison of the retrained
ResNet-18 student with its teacher, the comparisons at labelled cutoffs, the labelled few-image cutoff, the prevalence
estimates, the teacher LoRA seeds 1--2 (amendment 2 of E1.7), teacher test-time augmentation and the throughput
measurements. The additional baselines
were added after a literature review, once all pre-specified results of Protocols R and P had been computed; they do
not change any pre-specified result.

\begin{table}[htbp]
\floatconts{tab:app-protocols}
{\caption{Protocols in the order they were frozen (UTC). Each protocol file states the families, the expected direction
of every contrast and what was already known when it was written.}}
{%
\scriptsize
\setlength{\tabcolsep}{4pt}
\begin{tabular}{@{}llp{7.2cm}@{}}
\toprule
Protocol & Frozen & Content \\
\midrule
E1.7 (multi-teacher) & 2026-11-25 08:42 & Virchow2, UNI, UNI2-h teachers and ResNet-18 students; 9 contrasts. Amendment 1 (10:44): ConvNeXt-T students with H-optimus-0 (1 contrast, E1.7a); amendment 2 (12:04): teacher LoRA seeds 1--2 (descriptive). \\
R (recalibration) & 2026-11-25 08:55 & TGR against label-free baselines (9 contrasts) and against AdaBN with 1, 2, 5 images (3 contrasts). \\
T (TENT) & 2026-11-25 10:40 & TENT against AdaBN and against retraining, KD and no-teacher students (4 contrasts). \\
P (prevalence shift) & 2026-11-25 18:39 & Resampling design, TGR against the source positive rate and the source cutoff at $\pi=0.2, 0.6$ (8 contrasts). \\
N (non-inferiority) & 2026-11-25 22:03 & Margin 0.02; ConvNeXt-T against H-optimus-0 (2 contrasts) and Virchow2 (superiority + 2 contrasts). \\
F and C & 2026-11-26 06:44 & Teacher-seed robustness (2 contrasts + key secondary) and the ConvNeXt-T no-teacher control (2 contrasts + wording rule). \\
\bottomrule
\end{tabular}
}
\end{table}

\begin{table}[htbp]
\floatconts{tab:app-tests}
{\caption{Every pre-specified contrast, in the order the protocols were written. R18 / CNX: ResNet-18 / ConvNeXt-T student. Differences are new-scanner F1 with 95\% paired-bootstrap intervals; $p$ and ``met'' as defined in Appendix~\ref{app:prereg}. $^\ast$Point estimate known when the protocol was written. NI: non-inferiority at the margin 0.02.}}
{%
\tiny
\renewcommand{\arraystretch}{0.92}
\setlength{\tabcolsep}{2.5pt}
\begin{tabular}{@{}l>{\raggedright\arraybackslash}p{5.0cm}ccp{1.8cm}>{\raggedright\arraybackslash}p{2.4cm}@{}}
\toprule
Family & Contrast & Difference [95\% CI] & Holm $p$ & Expected & Outcome \\
\midrule
E1.7 & R18, Virchow2: KD $-$ no-teacher & +0.010 [-0.009, +0.030] & 1.000 & small, n.s. & met \\
E1.7 & R18, Virchow2: KD+TGA $-$ KD & +0.060 [+0.043, +0.079] & $<$0.001 & $>0$, significant & met \\
E1.7 & R18, Virchow2: KD+TGA $-$ no-teacher adaptation & +0.012 [-0.001, +0.027] & 0.367 & 0 to +0.03 & met \\
E1.7 & R18, UNI: KD $-$ no-teacher & +0.005 [-0.018, +0.027] & 1.000 & small, n.s. & met \\
E1.7 & R18, UNI: KD+TGA $-$ KD & +0.066 [+0.046, +0.091] & $<$0.001 & $>0$, significant & met \\
E1.7 & R18, UNI: KD+TGA $-$ no-teacher adaptation & +0.013 [+0.001, +0.027] & 0.227 & 0 to +0.03 & met \\
E1.7 & R18, UNI2-h: KD $-$ no-teacher & -0.007 [-0.031, +0.015] & 1.000 & small, n.s. & met \\
E1.7 & R18, UNI2-h: KD+TGA $-$ KD & +0.067 [+0.044, +0.094] & $<$0.001 & $>0$, significant & met \\
E1.7 & R18, UNI2-h: KD+TGA $-$ no-teacher adaptation & +0.002 [-0.014, +0.019] & 1.000 & 0 to +0.03 & met \\
E1.7a & CNX, H-optimus-0: KD+TGA $-$ KD & +0.040 [+0.026, +0.056] & $<$0.001 & $>0$, significant & met \\
R1 & R18 KD: TGR $-$ source cutoff$^\ast$ & +0.037 [+0.022, +0.053] & $<$0.001 & $>0$ & met \\
R1 & R18 KD: TGR $-$ source positive rate & +0.002 [-0.008, +0.012] & 0.675 & $>0$ & not met (positive, n.s.) \\
R1 & R18 KD: TGR $-$ EM prior & +0.210 [+0.176, +0.240] & $<$0.001 & $>0$ & met \\
R1 & R18 KD: TGR $-$ teacher pseudo-label cutoff & +0.004 [-0.001, +0.011] & 0.400 & none & -- \\
R1 & R18 KD: TGR $-$ AdaBN$^\ast$ & -0.010 [-0.024, +0.004] & 0.400 & $<0$ & not met (negative, n.s.) \\
R1 & CNX KD: TGR $-$ source cutoff & +0.036 [+0.022, +0.052] & $<$0.001 & $>0$ & met \\
R1 & CNX KD: TGR $-$ source positive rate & +0.019 [+0.009, +0.031] & $<$0.001 & $>0$ & met \\
R1 & CNX KD: TGR $-$ EM prior & +0.129 [+0.098, +0.161] & $<$0.001 & $>0$ & met \\
R1 & CNX KD: TGR $-$ teacher pseudo-label cutoff & +0.006 [+0.002, +0.012] & 0.022 & none & -- \\
R2 & R18 KD, 1 image: TGR $-$ AdaBN & +0.025 [+0.010, +0.041] & 0.004 & $\ge 0$ & met \\
R2 & R18 KD, 2 images: TGR $-$ AdaBN & +0.013 [-0.002, +0.027] & 0.164 & $\ge 0$ & met \\
R2 & R18 KD, 5 images: TGR $-$ AdaBN & +0.000 [-0.015, +0.015] & 0.958 & none & -- \\
T & R18 KD: TENT $-$ AdaBN & +0.005 [+0.002, +0.009] & 0.007 & within $\pm$0.01, n.s. & partly (within $\pm$0.01, but significant) \\
T & R18 KD: TENT $-$ KD+TGA & -0.012 [-0.025, +0.001] & 0.060 & $<0$, significant & not met (negative, n.s.) \\
T & R18 no-teacher: TENT $-$ AdaBN & +0.005 [+0.002, +0.008] & 0.010 & within $\pm$0.01, n.s. & partly (within $\pm$0.01, but significant) \\
T & R18 no-teacher: TENT $-$ no-teacher adaptation & -0.017 [-0.028, -0.006] & 0.010 & $<0$, significant & met \\
P & R18 KD, $\pi=0.2$: TGR $-$ source positive rate & +0.080 [+0.064, +0.096] & $<$0.001 & $>0$, significant & met \\
P & R18 KD, $\pi=0.6$: TGR $-$ source positive rate & +0.061 [+0.048, +0.077] & $<$0.001 & $>0$, significant & met \\
P & R18 KD, $\pi=0.2$: TGR $-$ source cutoff & +0.038 [+0.020, +0.058] & $<$0.001 & $>0$, significant & met \\
P & R18 KD, $\pi=0.6$: TGR $-$ source cutoff & +0.048 [+0.034, +0.064] & $<$0.001 & $>0$, significant & met \\
P & CNX KD, $\pi=0.2$: TGR $-$ source positive rate & +0.078 [+0.064, +0.094] & $<$0.001 & $>0$, significant & met \\
P & CNX KD, $\pi=0.6$: TGR $-$ source positive rate & +0.083 [+0.067, +0.103] & $<$0.001 & $>0$, significant & met \\
P & CNX KD, $\pi=0.2$: TGR $-$ source cutoff & +0.014 [-0.002, +0.031] & 0.080 & $>0$, significant & not met \\
P & CNX KD, $\pi=0.6$: TGR $-$ source cutoff & +0.051 [+0.038, +0.067] & $<$0.001 & $>0$, significant & met \\
N1a & CNX vs H-optimus-0 teacher: KD+TGA $-$ teacher$^\ast$ & -0.009 [-0.022, +0.003] & 0.085 & borderline & NI not established \\
N1b & CNX vs H-optimus-0 teacher: KD+TGR $-$ teacher$^\ast$ & -0.013 [-0.026, -0.001] & 0.151 & not non-inferior & NI not established (met) \\
N2c & CNX, Virchow2: KD+TGA $-$ KD (superiority) & +0.044 [+0.031, +0.059] & $<$0.001 & $>0$, significant & met \\
N2d & CNX vs Virchow2 teacher: KD+TGA $-$ teacher & -0.001 [-0.014, +0.012] & 0.004 & same as N1a (borderline) & non-inferior (better than expected) \\
N2e & CNX vs Virchow2 teacher: KD+TGR $-$ teacher & -0.016 [-0.031, -0.002] & 0.283 & not non-inferior & NI not established (met) \\
F1 & CNX vs H-optimus-0 LoRA seed 1: KD+TGA $-$ teacher & -0.012 [-0.024, +0.001] & 0.090 & NI not established & NI not established (met) \\
F2 & CNX vs H-optimus-0 LoRA seed 2: KD+TGA $-$ teacher & -0.005 [-0.018, +0.008] & 0.027 & non-inferior & NI not established (not met) \\
F key & CNX vs H-optimus-0, mean over LoRA seeds 0--2: KD+TGA $-$ teacher & -0.009 [-0.0198, +0.002] & 0.023 & borderline & non-inferior; partly seen data \\
C1 & CNX, H-optimus-0: KD+TGA $-$ no-teacher adaptation & +0.030 [+0.016, +0.045] & $<$0.001 & 0 to +0.04 & met \\
C2 & CNX, Virchow2: KD+TGA $-$ no-teacher adaptation & +0.030 [+0.017, +0.046] & $<$0.001 & 0 to +0.04 & met \\
\bottomrule
\end{tabular}
}
\end{table}

\FloatBarrier
\section{Implementation details}
\label{app:impl}

\begin{table}[t]
\floatconts{tab:ladder}
{\caption{The repair ladder. Each step acts on one layer of the student and needs different things on the unseen
scanner. All label-free steps are fitted on the unlabelled cells of one half of the new-scanner images and scored on the
other half, then the halves are swapped (cross-fitting).}}
{%
\scriptsize
\setlength{\tabcolsep}{3.5pt}
\begin{tabular}{@{}llcccl@{}}
\toprule
 & & \multicolumn{3}{c}{Needs on the new scanner} & \\
\cmidrule(lr){3-5}
Step & Layer & Unlabelled images & Teacher & Retraining & Students \\
\midrule
No-teacher student (0\%) & -- & -- & -- & -- & both \\
KD, source cutoff & -- & -- & -- & -- & both \\
\quad + TGR & operating point & \checkmark\ (about 5) & probabilities & -- & both \\
\quad + AdaBN & BN statistics & \checkmark & -- & -- & ResNet-18 \\
\quad + TENT & BN scale and shift & \checkmark & -- & at test time & ResNet-18 \\
KD + TGA & representation & \checkmark\ (25 per half) & soft labels & \checkmark & both \\
No-teacher adaptation & representation & \checkmark\ (25 per half) & -- & \checkmark & both \\
Labelled cutoff (upper bound) & operating point & labelled & -- & -- & both \\
Teacher (100\%) & -- & -- & -- & -- & -- \\
\bottomrule
\end{tabular}
}
\end{table}

\paragraph{Data.} Patches are cut around every annotated cell of the MIDOG++ breast subset at the native resolution:
$144\times144$ pixels for the students (random $128\times128$ crop during training, centre crop at test time) and
$256\times256$ pixels for the teachers (centre $224\times224$ crop). Per held-out scanner, the new-scanner set has 1{,}612
(CS2), 1{,}648 (S360) and 1{,}175 (XR) cells. The source validation set consists of eight fixed training images per
source scanner, drawn once with a fixed seed.

\paragraph{Teachers.} Frozen encoders in bfloat16 with LoRA (rank 16, $\alpha=16$, dropout 0.1) on the query--key--value,
output, and both MLP projections of every block; a linear head on [class token, mean patch token]; binary cross-entropy
with positive-class weighting; AdamW (learning rate $10^{-4}$, weight decay $10^{-4}$, one-cycle schedule), batch 16,
5 epochs, dihedral augmentation; the epoch and cutoff with the best source-validation F1 are kept. Normalisation as on
each model card.

\paragraph{Distillation.} For every source cell, the teacher's logit and its feature vector (class token and mean patch
token) are cached for 16 views: the 8 dihedral transforms, each once in the original colours and once with a random
H\&E stain jitter (per-channel scale $1\pm0.15$, shift $\pm0.03$ in the HED space), whose parameters are stored so that
the student sees exactly the same view. Student loss for KD: $T^2\ell(z/T, z_T/T) + 1-\cos(Wf, f_T)$, where $\ell$ is the binary cross-entropy between
the sigmoids of its arguments, $T=2$, $z$ and $z_T$ are the student and teacher logits, $f$ and $f_T$ their features, and
$W$ is a learned linear projection; no expert labels are used. No-teacher student: positive-weighted
BCE with the expert labels on the same views. AdamW (learning rate $3\times10^{-4}$ for ResNet-18, $10^{-4}$ for
ConvNeXt-T; weight decay $10^{-4}$; one-cycle schedule), batch 64, 60 epochs, checkpoint and cutoff by source-validation F1.

\paragraph{TGA and the no-teacher adaptation.} TGA adds the unlabelled cells of the fit half of the new scanner, with
the teacher's cached logits and features on the same 16 views, to the KD training set; nothing else changes. The
no-teacher adaptation keeps the expert labels on source cells and gives the fit-half cells the temperature-softened mean
logit of the ten no-teacher ResNet-18 students after AdaBN on that half, with no feature term. For each student seed and
held-out scanner, two models are trained (fit half A, fit half B), and each scores the other half.

\paragraph{Post-hoc steps.} TGR as in \equationref{eq:tgr}, with the bias found by bounded scalar minimisation in
$[-10, 10]$. AdaBN resets every BatchNorm layer's running statistics and re-estimates them as a cumulative average over
the fit-half cells. TENT trains only the BatchNorm affine parameters with Adam on the mean binary entropy of the fit-half
cells (batch 64, no augmentation), then re-estimates the statistics.

\paragraph{AdaBN + TGR departs from strict cross-fitting.} For this arm (\tableref{tab:main}, \figureref{fig:gap},
\tableref{tab:app-orders}), the offset applied to one half was fitted on the other half's cells as scored by the model
whose statistics were re-estimated on the scored half, not by the model that scores it. The scored half's unlabelled
images therefore enter its own offset through the BatchNorm statistics. Strictly cross-fitted predictions exist only
for H-optimus-0 (from the TENT runs); there the strict version gives 0.779 instead of 0.777 (S360: 0.762 instead of
0.755; CS2 and XR within 0.001), so the departure does not inflate this arm. For the other teachers the adapted
models' predictions on their own fit half were not saved.

\paragraph{Baselines for the operating point.} Source positive rate: the cutoff on the fit half that reproduces the
student's positive rate on source validation. UPA \citep{roschewitz2023automatic}: piecewise-linear quantile mapping
(101 quantiles) of fit-half scores onto source-validation scores, then the source cutoff. One bias, source mean: the
offset that makes the mean fit-half probability equal the mean source-validation probability. EM
\citep{saerens2002adjusting}, optionally after bias-corrected temperature scaling fitted on labelled source-validation
cells \citep{alexandari2020maximum}, and BBSE \citep{lipton2018detecting} estimate the target prevalence and shift the
logit accordingly. Teacher-referenced variants fit bias and temperature, map quantiles onto the teacher's logits,
reproduce the teacher's positive rate, or choose the F1-optimal cutoff against the teacher's hard labels.

\paragraph{Prevalence shift.} For each held-out scanner and target prevalence $\pi\in\{0.2,\dots,0.6\}$, 20 subsets are
drawn by removing randomly chosen positives (if $\pi$ is below the natural prevalence) or negatives (otherwise); every
method is refitted on each subset with the same cross-fitting, and counts are pooled over the 20 subsets.

\paragraph{Software and hardware.} PyTorch and timm on one NVIDIA RTX 3090 Ti (24 GB). Foundation-model weights were
obtained from their official repositories under their licences.

\FloatBarrier
\section{Additional results}

\FloatBarrier
\subsection{Per-scanner results}
\label{app:perscanner}
\begin{table}[htbp]
\floatconts{tab:app-perscanner}
{\caption{New-scanner F1 per held-out scanner (mean over 10 student seeds) and mean over scanners. The no-teacher rows of the ResNet-18 student do not depend on the teacher and are shown once.}}
{%
\scriptsize
\setlength{\tabcolsep}{3pt}
\begin{tabular}{@{}lcccc@{}}
\toprule
Arm & CS2 & S360 & XR & Mean \\
\midrule
\multicolumn{5}{@{}l}{\textit{ResNet-18, H-optimus-0 teacher}} \\
\quad No-teacher student & 0.754 & 0.635 & 0.745 & 0.711 \\
\quad KD & 0.798 & 0.639 & 0.738 & 0.725 \\
\quad KD + TGR & 0.798 & 0.717 & 0.770 & 0.762 \\
\quad KD + AdaBN + TGR & 0.787 & 0.755 & 0.788 & 0.777 \\
\quad KD + TGA & 0.796 & 0.779 & 0.790 & 0.788 \\
\quad No-teacher adaptation & 0.759 & 0.773 & 0.776 & 0.769 \\
\quad Teacher (seed 0) & 0.833 & 0.806 & 0.803 & 0.814 \\
\multicolumn{5}{@{}l}{\textit{ResNet-18, Virchow2 teacher}} \\
\quad KD & 0.791 & 0.641 & 0.733 & 0.722 \\
\quad KD + TGR & 0.790 & 0.715 & 0.738 & 0.748 \\
\quad KD + AdaBN + TGR & 0.790 & 0.770 & 0.764 & 0.775 \\
\quad KD + TGA & 0.797 & 0.774 & 0.774 & 0.781 \\
\quad Teacher (seed 0) & 0.828 & 0.801 & 0.788 & 0.805 \\
\multicolumn{5}{@{}l}{\textit{ResNet-18, UNI2-h teacher}} \\
\quad KD & 0.776 & 0.606 & 0.731 & 0.704 \\
\quad KD + TGR & 0.781 & 0.681 & 0.769 & 0.744 \\
\quad KD + AdaBN + TGR & 0.779 & 0.742 & 0.784 & 0.768 \\
\quad KD + TGA & 0.775 & 0.756 & 0.783 & 0.771 \\
\quad Teacher (seed 0) & 0.819 & 0.779 & 0.791 & 0.797 \\
\multicolumn{5}{@{}l}{\textit{ResNet-18, UNI teacher}} \\
\quad KD & 0.778 & 0.614 & 0.757 & 0.716 \\
\quad KD + TGR & 0.748 & 0.676 & 0.737 & 0.720 \\
\quad KD + AdaBN + TGR & 0.766 & 0.699 & 0.749 & 0.738 \\
\quad KD + TGA & 0.799 & 0.765 & 0.783 & 0.782 \\
\quad Teacher (seed 0) & 0.786 & 0.756 & 0.758 & 0.767 \\
\multicolumn{5}{@{}l}{\textit{ConvNeXt-T, H-optimus-0 teacher}} \\
\quad No-teacher student & 0.780 & 0.684 & 0.743 & 0.736 \\
\quad KD & 0.813 & 0.714 & 0.768 & 0.765 \\
\quad KD + TGR & 0.828 & 0.785 & 0.788 & 0.801 \\
\quad KD + TGA & 0.803 & 0.803 & 0.809 & 0.805 \\
\quad No-teacher adaptation & 0.759 & 0.777 & 0.789 & 0.775 \\
\quad Teacher (seed 0) & 0.833 & 0.806 & 0.803 & 0.814 \\
\multicolumn{5}{@{}l}{\textit{ConvNeXt-T, Virchow2 teacher}} \\
\quad No-teacher student & 0.780 & 0.684 & 0.743 & 0.736 \\
\quad KD & 0.826 & 0.701 & 0.756 & 0.761 \\
\quad KD + TGR & 0.828 & 0.781 & 0.760 & 0.789 \\
\quad KD + TGA & 0.826 & 0.789 & 0.800 & 0.805 \\
\quad No-teacher adaptation & 0.759 & 0.777 & 0.789 & 0.775 \\
\quad Teacher (seed 0) & 0.828 & 0.801 & 0.788 & 0.805 \\
\bottomrule
\end{tabular}
}
\end{table}

\FloatBarrier
\subsection{The ladder in both orders, the statistics swap, and the UNI teacher}
\label{app:ladder}

\tableref{tab:app-orders} shows why we report Shapley credit: AdaBN alone already moves the operating point, so the
gain of TGR is smaller when it comes second. \tableref{tab:app-bnswap} shows that the retrained (TGA) students' gain on
S360 and XR is not carried by their BatchNorm statistics. The seed-0 UNI teacher chose source cutoffs of 0.25 to 0.28
(seeds 1 and 2: 0.39 to 0.62) and over-calls mitoses on the new scanners; TGR transfers this operating point to the
student (threshold layer $-10\%$ of a small advantage), whereas retraining does not depend on it, so that the retrained
student exceeds this particular teacher ($+0.016$, $+0.004$ to $+0.029$).

\begin{table}[htbp]
\floatconts{tab:app-orders}
{\caption{The two post-hoc layers of the ResNet-18 KD student in both orders (F1 gain of each step) and their order-free Shapley credit with 95\% intervals. AdaBN also moves the operating point, so the threshold gain shrinks when AdaBN comes first.}}
{%
\scriptsize
\setlength{\tabcolsep}{3pt}
\begin{tabular}{@{}lcccccc@{}}
\toprule
Teacher & \makecell{TGR\\first} & \makecell{then\\AdaBN} & \makecell{AdaBN\\first} & \makecell{then\\TGR} & Shapley: threshold & Shapley: statistics \\
\midrule
H-optimus-0 & +0.037 & +0.015 & +0.047 & +0.005 & +0.021 [+0.013, +0.030] & +0.031 [+0.015, +0.049] \\
Virchow2 & +0.026 & +0.027 & +0.047 & +0.005 & +0.016 [+0.008, +0.024] & +0.037 [+0.021, +0.054] \\
UNI2-h & +0.039 & +0.025 & +0.057 & +0.007 & +0.023 [+0.016, +0.032] & +0.041 [+0.023, +0.061] \\
UNI & +0.004 & +0.017 & +0.036 & -0.015 & -0.005 [-0.019, +0.009] & +0.027 [+0.010, +0.044] \\
\bottomrule
\end{tabular}
}
\end{table}

\begin{table}[htbp]
\floatconts{tab:app-bnswap}
{\caption{ResNet-18 students retrained on the new scanner's fit half, evaluated with their own BatchNorm statistics and with statistics re-estimated on source cells only (cutoff re-chosen on source validation). Drop: mean over 10 seeds, interval from a seed bootstrap. On S360 and XR, where retraining helps, the statistics explain little of its gain; on CS2 weights and statistics have co-adapted, so the swap is not interpretable as the statistics' contribution.}}
{%
\scriptsize
\setlength{\tabcolsep}{3pt}
\begin{tabular}{@{}llccc@{}}
\toprule
Student & Scanner & Own statistics & Source statistics & Drop [95\% CI] \\
\midrule
No-teacher adaptation & CS2 & 0.759 & 0.739 & 0.020 [0.014, 0.027] \\
No-teacher adaptation & S360 & 0.773 & 0.769 & 0.004 [-0.001, 0.009] \\
No-teacher adaptation & XR & 0.776 & 0.759 & 0.017 [0.005, 0.027] \\
KD + TGA & CS2 & 0.796 & 0.667 & 0.128 [0.089, 0.170] \\
KD + TGA & S360 & 0.779 & 0.769 & 0.011 [0.002, 0.019] \\
KD + TGA & XR & 0.790 & 0.783 & 0.007 [0.002, 0.013] \\
\bottomrule
\end{tabular}
}
\end{table}

\FloatBarrier
\subsection{Recalibration baselines}
\label{app:baselines}
\begin{table}[t]
\floatconts{tab:recal}
{\caption{Recalibrating the operating point without new-scanner labels (KD students of H-optimus-0). F1 on the unseen
scanner at its natural mitosis prevalence and after resampling its cells to a prevalence of 0.2 or 0.6 (source scanners:
0.32 to 0.45). Bold: best label-free method in the column. $^\ddagger$Also uses the labels of the source-validation
cells. At a single operating point, UPA reduces to keeping the source positive rate. Paired differences to TGR with 95\%
intervals are in \tableref{tab:recal-diff}.}}
{%
\scriptsize
\setlength{\tabcolsep}{4pt}
\begin{tabular}{@{}lccc@{\hspace{10pt}}ccc@{}}
\toprule
 & \multicolumn{3}{c}{ResNet-18 KD student} & \multicolumn{3}{c}{ConvNeXt-T KD student} \\
\cmidrule(lr){2-4}\cmidrule(l){5-7}
Prevalence on the new scanner & natural & 0.2 & 0.6 & natural & 0.2 & 0.6 \\
\midrule
\multicolumn{7}{@{}l}{\textit{No recalibration}} \\
\quad Source cutoff & 0.725 & 0.609 & 0.784 & 0.765 & 0.661 & 0.815 \\
\addlinespace[2pt]
\multicolumn{7}{@{}l}{\textit{Teacher-free (reference: the student's source-validation outputs)}} \\
\quad UPA, quantile alignment \citep{roschewitz2023automatic} & 0.760 & 0.566 & 0.771 & 0.781 & 0.597 & 0.783 \\
\quad Source positive rate & 0.760 & 0.566 & 0.770 & 0.782 & 0.597 & 0.784 \\
\quad One bias, source mean probability & 0.760 & 0.601 & 0.765 & 0.779 & 0.650 & 0.741 \\
\quad EM prior adjustment & 0.552 & 0.411 & 0.705 & 0.671 & 0.489 & 0.793 \\
\quad EM + bias-corrected temperature scaling$^\ddagger$ & 0.521 & 0.407 & 0.681 & 0.678 & 0.497 & 0.794 \\
\quad BBSE$^\ddagger$ & 0.545 & 0.410 & 0.720 & 0.674 & 0.460 & 0.797 \\
\addlinespace[2pt]
\multicolumn{7}{@{}l}{\textit{Teacher-referenced (reference: the teacher on the same unlabelled cells)}} \\
\quad TGR: one bias (ours) & 0.762 & 0.646 & 0.832 & \textbf{0.801} & 0.675 & 0.866 \\
\quad Bias + temperature & 0.760 & \textbf{0.649} & 0.830 & 0.799 & \textbf{0.684} & 0.866 \\
\quad Quantile map onto the teacher & \textbf{0.763} & 0.632 & \textbf{0.839} & 0.799 & 0.669 & 0.867 \\
\quad Teacher positive rate & 0.762 & 0.632 & \textbf{0.839} & 0.799 & 0.668 & 0.867 \\
\quad Cutoff from teacher pseudo-labels & 0.757 & 0.618 & \textbf{0.839} & 0.794 & 0.658 & \textbf{0.870} \\
\addlinespace[2pt]
\multicolumn{7}{@{}l}{\textit{Uses new-scanner labels}} \\
\quad Labelled cutoff (upper bound) & 0.762 & 0.666 & 0.842 & 0.800 & 0.705 & 0.872 \\
\addlinespace[2pt]
\multicolumn{7}{@{}l}{\textit{Reference}} \\
\quad Teacher alone & 0.814 & 0.681 & 0.883 & 0.814 & 0.681 & 0.883 \\
\bottomrule
\end{tabular}
}
\end{table}

\begin{table}[htbp]
\floatconts{tab:recal-diff}
{\caption{TGR minus each recalibration method (F1, 95\% paired-bootstrap interval; descriptive). A separate bootstrap run from \tableref{tab:app-tests}, so shared contrasts can differ in the third decimal of an interval limit.}}
{%
\tiny
\setlength{\tabcolsep}{1.5pt}
\begin{tabular}{@{}lcccccc@{}}
\toprule
Method & \multicolumn{3}{c}{ResNet-18 KD: natural, 0.2, 0.6} & \multicolumn{3}{c}{ConvNeXt-T KD: natural, 0.2, 0.6} \\
\midrule
Source cutoff & \makecell{+0.037\\[-1pt]{\tiny[+0.022, +0.053]}} & \makecell{+0.038\\[-1pt]{\tiny[+0.020, +0.058]}} & \makecell{+0.048\\[-1pt]{\tiny[+0.034, +0.064]}} & \makecell{+0.036\\[-1pt]{\tiny[+0.022, +0.052]}} & \makecell{+0.014\\[-1pt]{\tiny[-0.002, +0.031]}} & \makecell{+0.051\\[-1pt]{\tiny[+0.038, +0.067]}} \\
UPA & \makecell{+0.002\\[-1pt]{\tiny[-0.008, +0.012]}} & \makecell{+0.081\\[-1pt]{\tiny[+0.065, +0.097]}} & \makecell{+0.061\\[-1pt]{\tiny[+0.047, +0.077]}} & \makecell{+0.020\\[-1pt]{\tiny[+0.010, +0.032]}} & \makecell{+0.078\\[-1pt]{\tiny[+0.064, +0.094]}} & \makecell{+0.084\\[-1pt]{\tiny[+0.067, +0.104]}} \\
Source positive rate & \makecell{+0.002\\[-1pt]{\tiny[-0.008, +0.012]}} & \makecell{+0.080\\[-1pt]{\tiny[+0.064, +0.096]}} & \makecell{+0.061\\[-1pt]{\tiny[+0.048, +0.077]}} & \makecell{+0.019\\[-1pt]{\tiny[+0.009, +0.031]}} & \makecell{+0.078\\[-1pt]{\tiny[+0.064, +0.094]}} & \makecell{+0.083\\[-1pt]{\tiny[+0.067, +0.103]}} \\
One bias, source mean & \makecell{+0.001\\[-1pt]{\tiny[-0.007, +0.011]}} & \makecell{+0.046\\[-1pt]{\tiny[+0.031, +0.061]}} & \makecell{+0.067\\[-1pt]{\tiny[+0.050, +0.087]}} & \makecell{+0.022\\[-1pt]{\tiny[+0.010, +0.037]}} & \makecell{+0.025\\[-1pt]{\tiny[+0.016, +0.034]}} & \makecell{+0.126\\[-1pt]{\tiny[+0.096, +0.158]}} \\
EM prior & \makecell{+0.210\\[-1pt]{\tiny[+0.175, +0.240]}} & \makecell{+0.235\\[-1pt]{\tiny[+0.213, +0.257]}} & \makecell{+0.127\\[-1pt]{\tiny[+0.097, +0.158]}} & \makecell{+0.129\\[-1pt]{\tiny[+0.098, +0.161]}} & \makecell{+0.186\\[-1pt]{\tiny[+0.152, +0.216]}} & \makecell{+0.073\\[-1pt]{\tiny[+0.054, +0.095]}} \\
EM + BCTS & \makecell{+0.241\\[-1pt]{\tiny[+0.212, +0.264]}} & \makecell{+0.240\\[-1pt]{\tiny[+0.220, +0.261]}} & \makecell{+0.150\\[-1pt]{\tiny[+0.117, +0.183]}} & \makecell{+0.122\\[-1pt]{\tiny[+0.091, +0.155]}} & \makecell{+0.178\\[-1pt]{\tiny[+0.141, +0.212]}} & \makecell{+0.073\\[-1pt]{\tiny[+0.053, +0.095]}} \\
BBSE & \makecell{+0.216\\[-1pt]{\tiny[+0.180, +0.245]}} & \makecell{+0.236\\[-1pt]{\tiny[+0.213, +0.260]}} & \makecell{+0.112\\[-1pt]{\tiny[+0.084, +0.140]}} & \makecell{+0.127\\[-1pt]{\tiny[+0.096, +0.158]}} & \makecell{+0.215\\[-1pt]{\tiny[+0.177, +0.248]}} & \makecell{+0.070\\[-1pt]{\tiny[+0.050, +0.092]}} \\
Bias + temp. (teacher) & \makecell{+0.002\\[-1pt]{\tiny[+0.000, +0.004]}} & \makecell{-0.002\\[-1pt]{\tiny[-0.007, +0.002]}} & \makecell{+0.001\\[-1pt]{\tiny[+0.001, +0.002]}} & \makecell{+0.001\\[-1pt]{\tiny[-0.001, +0.004]}} & \makecell{-0.009\\[-1pt]{\tiny[-0.013, -0.004]}} & \makecell{+0.001\\[-1pt]{\tiny[-0.000, +0.002]}} \\
Quantile map (teacher) & \makecell{-0.001\\[-1pt]{\tiny[-0.006, +0.004]}} & \makecell{+0.014\\[-1pt]{\tiny[+0.008, +0.021]}} & \makecell{-0.007\\[-1pt]{\tiny[-0.011, -0.004]}} & \makecell{+0.002\\[-1pt]{\tiny[-0.002, +0.006]}} & \makecell{+0.007\\[-1pt]{\tiny[+0.002, +0.013]}} & \makecell{-0.001\\[-1pt]{\tiny[-0.004, +0.003]}} \\
Teacher positive rate & \makecell{-0.001\\[-1pt]{\tiny[-0.006, +0.005]}} & \makecell{+0.014\\[-1pt]{\tiny[+0.008, +0.021]}} & \makecell{-0.007\\[-1pt]{\tiny[-0.010, -0.003]}} & \makecell{+0.002\\[-1pt]{\tiny[-0.002, +0.006]}} & \makecell{+0.007\\[-1pt]{\tiny[+0.002, +0.013]}} & \makecell{-0.000\\[-1pt]{\tiny[-0.003, +0.003]}} \\
Teacher pseudo-label cut. & \makecell{+0.004\\[-1pt]{\tiny[-0.001, +0.011]}} & \makecell{+0.028\\[-1pt]{\tiny[+0.020, +0.037]}} & \makecell{-0.007\\[-1pt]{\tiny[-0.012, -0.003]}} & \makecell{+0.006\\[-1pt]{\tiny[+0.002, +0.012]}} & \makecell{+0.018\\[-1pt]{\tiny[+0.012, +0.025]}} & \makecell{-0.003\\[-1pt]{\tiny[-0.007, +0.001]}} \\
Labelled cutoff & \makecell{-0.001\\[-1pt]{\tiny[-0.007, +0.006]}} & \makecell{-0.020\\[-1pt]{\tiny[-0.032, -0.008]}} & \makecell{-0.010\\[-1pt]{\tiny[-0.016, -0.004]}} & \makecell{+0.000\\[-1pt]{\tiny[-0.006, +0.008]}} & \makecell{-0.030\\[-1pt]{\tiny[-0.046, -0.012]}} & \makecell{-0.006\\[-1pt]{\tiny[-0.011, -0.001]}} \\
Teacher alone & \makecell{-0.052\\[-1pt]{\tiny[-0.069, -0.039]}} & \makecell{-0.035\\[-1pt]{\tiny[-0.056, -0.017]}} & \makecell{-0.052\\[-1pt]{\tiny[-0.066, -0.039]}} & \makecell{-0.013\\[-1pt]{\tiny[-0.026, -0.001]}} & \makecell{-0.006\\[-1pt]{\tiny[-0.025, +0.011]}} & \makecell{-0.017\\[-1pt]{\tiny[-0.027, -0.007]}} \\
\bottomrule
\end{tabular}
}
\end{table}

\begin{table}[htbp]
\floatconts{tab:app-priors}
{\caption{Label-free estimates of the new scanner's mitosis prevalence (mean over 10 student seeds) against the truth. The prior-shift estimators follow the student's own shifted positive rate and amplify it (S360, XR).}}
{%
\scriptsize
\setlength{\tabcolsep}{3pt}
\begin{tabular}{@{}llcccccc@{}}
\toprule
KD student & Scanner & True & \makecell{Student positive\\rate} & EM & EM + BCTS & BBSE & \makecell{Teacher positive\\rate} \\
\midrule
ConvNeXt-T & CS2 & 0.43 & 0.39 & 0.53 & 0.40 & 0.37 & 0.43 \\
ConvNeXt-T & S360 & 0.35 & 0.24 & 0.20 & 0.15 & 0.17 & 0.47 \\
ConvNeXt-T & XR & 0.38 & 0.55 & 0.62 & 0.59 & 0.55 & 0.47 \\
ResNet-18 & CS2 & 0.43 & 0.43 & 0.58 & 0.45 & 0.41 & 0.43 \\
ResNet-18 & S360 & 0.35 & 0.21 & 0.09 & 0.04 & 0.08 & 0.47 \\
ResNet-18 & XR & 0.38 & 0.58 & 0.66 & 0.67 & 0.61 & 0.47 \\
\bottomrule
\end{tabular}
}
\end{table}

\FloatBarrier
\subsection{Few images}
\label{app:few}
\begin{table}[htbp]
\floatconts{tab:app-budget}
{\caption{New-scanner F1 with $n$ images per half from the new scanner (mean over folds, seeds and 10 image draws; the same draws for every method). TGR and AdaBN use the images unlabelled; the labelled cutoff uses their labels. Source cutoff: 0.725 (ResNet-18), 0.765 (ConvNeXt-T).}}
{%
\footnotesize
\setlength{\tabcolsep}{5pt}
\begin{tabular}{@{}cccccc@{}}
\toprule
$n$ & R18: TGR & R18: AdaBN & R18: labelled cutoff & CNX: TGR & CNX: labelled cutoff \\
\midrule
1 & 0.740 & 0.715 & 0.646 & 0.791 & 0.705 \\
2 & 0.755 & 0.743 & 0.734 & 0.798 & 0.775 \\
5 & 0.757 & 0.757 & 0.743 & 0.799 & 0.789 \\
10 & 0.762 & 0.772 & 0.760 & 0.801 & 0.798 \\
25 & 0.762 & 0.772 & 0.762 & 0.801 & 0.800 \\
\bottomrule
\end{tabular}
}
\end{table}

\FloatBarrier
\subsection{Teachers and inference cost}
\label{app:teachers}
\begin{table}[htbp]
\floatconts{tab:app-teachers}
{\caption{Teachers on the unseen scanner: F1 for three LoRA seeds (mean over the three held-out scanners), their mean, and the range of AUROC over seeds; below, seed 0 with eight-view test-time augmentation and a TTA cutoff from source validation. Seed-0 values are re-evaluated from the saved LoRA weights and can differ in the third decimal from the cached predictions used elsewhere (UNI: 0.766 here, 0.767 in \tableref{tab:main}).}}
{%
\footnotesize
\setlength{\tabcolsep}{5pt}
\begin{tabular}{@{}lccccc@{}}
\toprule
Teacher & Seed 0 & Seed 1 & Seed 2 & Mean & AUROC range \\
\midrule
H-optimus-0 & 0.814 & 0.821 & 0.801 & 0.812 & 0.925--0.929 \\
Virchow2 & 0.805 & 0.805 & 0.803 & 0.804 & 0.917--0.918 \\
UNI2-h & 0.797 & 0.792 & 0.787 & 0.792 & 0.913--0.916 \\
UNI & 0.766 & 0.782 & 0.786 & 0.778 & 0.905--0.908 \\
\midrule
H-optimus-0, 8-view TTA (seed 0) & 0.812 & & & & \\
Virchow2, 8-view TTA (seed 0) & 0.797 & & & & \\
\bottomrule
\end{tabular}
}
\end{table}

\begin{table}[htbp]
\floatconts{tab:app-efficiency}
{\caption{Inference cost on one RTX 3090 Ti: random inputs at the resolution used in training, mixed precision (fp16 students, bf16 teachers), 5 warm-up batches, then 30 timed batches, median of 3 repetitions. Parameters include LoRA and heads.}}
{%
\footnotesize
\setlength{\tabcolsep}{5pt}
\begin{tabular}{@{}lrrrrr@{}}
\toprule
Model & Params (M) & Input (px) & Batch & Cells/s & Peak memory (GB) \\
\midrule
ResNet-18 student & 11.2 & 128 & 256 & 23,073 & 0.40 \\
ConvNeXt-T student & 27.8 & 128 & 256 & 7,899 & 0.72 \\
H-optimus-0 teacher & 1,150.5 & 224 & 64 & 80 & 3.29 \\
Virchow2 teacher & 641.7 & 224 & 64 & 128 & 2.15 \\
UNI2-h teacher & 690.8 & 224 & 64 & 132 & 2.41 \\
UNI teacher & 309.6 & 224 & 64 & 342 & 1.08 \\
\bottomrule
\end{tabular}
}
\end{table}

\FloatBarrier
\subsection{TENT and the scanner probe}
\label{app:tent}
\begin{table}[htbp]
\floatconts{tab:app-tent}
{\caption{TENT on the ResNet-18 students (new-scanner F1, mean over seeds and scanners). Primary setting: learning rate $10^{-3}$, one epoch; the others are a descriptive sensitivity analysis.}}
{%
\footnotesize
\setlength{\tabcolsep}{5pt}
\begin{tabular}{@{}lccccc@{}}
\toprule
Student & lr & Epochs & AdaBN & AdaBN + TENT & + TGR \\
\midrule
No-teacher & 0.0001 & 1 & 0.748 & 0.748 & 0.757 \\
No-teacher & 0.001 & 1 & 0.748 & 0.753 & 0.759 \\
No-teacher & 0.001 & 3 & 0.748 & 0.756 & 0.760 \\
No-teacher & 0.01 & 1 & 0.748 & 0.693 & 0.707 \\
KD & 0.0001 & 1 & 0.772 & 0.772 & 0.780 \\
KD & 0.001 & 1 & 0.772 & 0.777 & 0.784 \\
KD & 0.001 & 3 & 0.772 & 0.778 & 0.784 \\
KD & 0.01 & 1 & 0.772 & 0.726 & 0.735 \\
\bottomrule
\end{tabular}
}
\end{table}

\begin{table}[htbp]
\floatconts{tab:app-probe}
{\caption{Linear probe for the training scanner on source-cell features (image-grouped cross-validation, two classes, mean over folds and seeds). All models encode the scanner almost perfectly, so robustness does not mean that scanner information is absent.}}
{%
\footnotesize
\setlength{\tabcolsep}{5pt}
\begin{tabular}{@{}lcc@{}}
\toprule
Features & Dimension & Probe accuracy \\
\midrule
H-optimus-0 & 3072 & 1.000 \\
Virchow2 & 2560 & 0.999 \\
UNI2-h & 3072 & 1.000 \\
UNI & 2048 & 1.000 \\
H-optimus-0 (PCA 512) & 512 & 0.814 \\
Virchow2 (PCA 512) & 512 & 0.824 \\
UNI2-h (PCA 512) & 512 & 0.853 \\
UNI (PCA 512) & 512 & 0.859 \\
ResNet-18 no-teacher & 512 & 0.962 \\
ResNet-18 KD (H-optimus-0) & 512 & 0.999 \\
ResNet-18 KD + TGA (H-optimus-0, half A) & 512 & 0.999 \\
\bottomrule
\end{tabular}
}
\end{table}

\FloatBarrier
\section{Protocols F and C}
\label{app:fc}
Both protocols were frozen together (\tableref{tab:app-protocols}) before any of the new models they concern was
trained; the ConvNeXt-T KD + TGA students of C1 and C2 and the seed-0 term of F's key secondary estimate already existed.

\paragraph{Protocol C: no-teacher control for ConvNeXt-T.} The no-teacher adaptation recipe of the ResNet-18 control
was applied unchanged to ConvNeXt-T (60 students: three held-out scanners, two fit halves, ten seeds): expert labels on
the source cells plus the soft answers of the AdaBN-adapted ensemble of ten no-teacher ResNet-18 students on the
unlabelled fit half, scored on the other half. Both pre-specified contrasts (two-sided, Holm) are positive and inside
the expected range of 0 to +0.04: KD + TGA minus no-teacher adaptation is +0.030 (+0.016 to +0.045) with H-optimus-0
and +0.030 (+0.017 to +0.046) with Virchow2 (\tableref{tab:app-tests}). The protocol fixed the wording of the main
text in advance by the share of TGA's gain over the no-teacher student that the no-teacher adaptation reaches: if it
was at least 0.5 for both teachers, the text would say that for both students most of the retraining gain comes from
the unlabelled images. It is 0.57 (0.34 to 0.76) for both teachers, so this wording is used; its interval includes values
below 0.5. The adapted no-teacher ConvNeXt-T reaches 0.759 (CS2), 0.777 (S360) and 0.789 (XR), and stays below the KD
student with TGR ($-0.025$, $-0.040$ to $-0.013$, with H-optimus-0; $-0.014$, $-0.031$ to $-0.0002$, with Virchow2;
descriptive).

\paragraph{Protocol F: teacher LoRA seed.} The ConvNeXt-T students of H-optimus-0 were distilled again, with the recipe
unchanged, from LoRA seeds 1 and 2 of the teacher (180 students: for each teacher seed, KD, and KD + TGA on both fit halves, for ten seeds and
three held-out scanners). The teacher caches were rebuilt exactly as for seed 0, so the stain jitter of the cached views
also differs between teacher seeds, as in a complete re-run of the pipeline. Each student is compared with the
teacher training it was distilled from (one view, the teacher's own source cutoff). The teachers reach 0.814, 0.821 and
0.801 for seeds 0, 1 and 2, the retrained students 0.805, 0.809 and 0.796. The protocol expected the student to land near
0.80 to 0.81 whatever the teacher, so that non-inferiority would fail for seed 1 and hold for seed 2. The first
expectation was met and the second was not: seed 1, $-0.012$ ($-0.024$ to $+0.001$, Holm $p=0.090$); seed 2, $-0.005$
($-0.018$ to $+0.008$), whose interval lies within the margin but whose Holm-adjusted $p$ of 0.027 misses 0.025
(unadjusted 0.013). The pre-specified key secondary estimate, the gap averaged over the three trainings, is $-0.009$
($-0.0198$ to $+0.0023$, one-sided $p=0.023$); it meets the non-inferiority criterion, but only just, and its seed-0 term
was known when the protocol was written. Descriptively, the retrained student recovers 86\% (73\% to 101\%) and 92\%
(74\% to 114\%) of the advantage of the seed-1 and seed-2 teachers (seed 0: 88\%), KD alone 48\% and 63\% (seed 0:
37\%), and KD with TGR stays 0.017 ($-0.029$ to $-0.005$) and 0.007 ($-0.021$ to $+0.006$) below them. Across the three
trainings, the retrained student's F1 varies by 0.013 (0.006 to 0.021), the teacher's by 0.020 (0.010 to 0.030).

\FloatBarrier
\section{Compute, code and data}
\label{app:compute}

The study trained 936 networks (36 LoRA teachers and 900 students) in about 22 GPU hours of training-loop time on
one RTX 3090 Ti, except the three seed-0 H-optimus-0 teachers, which a co-author trained separately
(\tableref{tab:app-compute}); 150 further exploratory networks are not reported. MIDOG++ is publicly
available \citep{aubreville2023comprehensive}. Code, per-cell results and the pre-registration protocols are available at \url{https://github.com/AaronG211/scanner-robust-distillation}.

\begin{table}[htbp]
\floatconts{tab:app-compute}
{\caption{Networks trained for the study and their training-loop time on one RTX 3090 Ti (excluding teacher caching and evaluation). Not in the paper: two superseded exploratory variants.}}
{%
\tiny
\setlength{\tabcolsep}{3pt}
\begin{tabular}{@{}llrrc@{}}
\toprule
Group & Networks & $n$ & GPU hours & In paper \\
\midrule
teacher & H-optimus-0 LoRA, seed 0 & 3 & 1.24 & yes \\
teacher & uni-lora, seed 0 & 3 & 0.12 & yes \\
teacher & uni2h-lora, seed 0 & 3 & 0.27 & yes \\
teacher & virchow2-lora, seed 0 & 3 & 0.28 & yes \\
teacher & h-optimus-0-lora, seeds 1-2 & 6 & 1.29 & yes \\
teacher & uni-lora, seeds 1-2 & 6 & 0.24 & yes \\
teacher & uni2h-lora, seeds 1-2 & 6 & 0.54 & yes \\
teacher & virchow2-lora, seeds 1-2 & 6 & 0.56 & yes \\
ResNet-18 & no-teacher student & 30 & 0.26 & yes \\
ResNet-18 & KD, H-optimus-0 & 30 & 0.26 & yes \\
ResNet-18 & KD, Virchow2 / UNI / UNI2-h & 90 & 0.85 & yes \\
ResNet-18 & KD + TGA, H-optimus-0 (halves A/B) & 60 & 0.69 & yes \\
ResNet-18 & KD + TGA, Virchow2 / UNI / UNI2-h (halves A/B) & 180 & 2.31 & yes \\
ResNet-18 & no-teacher adaptation (halves A/B) & 60 & 0.73 & yes \\
ConvNeXt-T & no-teacher student & 30 & 0.66 & yes \\
ConvNeXt-T & KD, H-optimus-0 / Virchow2 & 60 & 1.31 & yes \\
ConvNeXt-T & KD + TGA, H-optimus-0 / Virchow2 (halves A/B) & 120 & 3.55 & yes \\
ConvNeXt-T & no-teacher adaptation (halves A/B; Protocol C) & 60 & 1.73 & yes \\
ConvNeXt-T & KD, H-optimus-0 LoRA seeds 1-2 (Protocol F) & 60 & 1.28 & yes \\
ConvNeXt-T & KD + TGA, H-optimus-0 LoRA seeds 1-2 (halves A/B; Protocol F) & 120 & 3.50 & yes \\
ResNet-18 & KD + target adaptation without cross-fitting (superseded) & 30 & 0.40 & no \\
ResNet-18 & TGA variants (ResNet pseudo-labeller, soft-only) & 120 & 1.48 & no \\
\bottomrule
\end{tabular}
}
\end{table}

\FloatBarrier

\end{document}